\documentclass[lettersize,journal]{IEEEtran}
\usepackage{enumitem}
\usepackage{amsfonts}
\usepackage{enumitem,kantlipsum}
\usepackage{textcomp}
\usepackage{stfloats}
\usepackage{url}
\usepackage{caption}
\usepackage{graphics}
\usepackage{tabularx}
\usepackage[dvipsnames,table]{xcolor}
\usepackage{verbatim}
\usepackage{pifont}
\usepackage{float}
\usepackage{amsmath}
\usepackage{cancel}
\usepackage{libertinust1math}
\usepackage{pgfplots}
\usepackage{amsmath}
\usepackage{forest}
\usepackage{graphicx}
\usepackage{cite}
\usepackage{etoolbox}
\usepackage{graphicx}
\usepackage{float}
\usepackage{lscape}
\usepackage{longtable}
\usepackage{ltxtable}
\usepackage{hhline}
\usepackage{caption}
\usepackage{wasysym}
\usepackage{subcaption}
\usepackage{algorithm} 
\usepackage{color} 
\usepackage{colortbl}
\usepackage{algpseudocode}
\usepackage{verbatim}
\usepackage{lipsum}
\usetikzlibrary{automata, arrows.meta, positioning}
\usetikzlibrary{decorations.pathreplacing}
\usepackage{amsthm}
\usepackage{booktabs}
\usepackage{amsmath}
\usepackage{algorithm}
\usepackage{inputenc}
\usepackage{adjustbox}
\usepackage{url} 
\usepackage{multirow}
\usepackage{color,soul}
\usepackage{lipsum}
\usepackage{mathtools}
\usepackage{cuted}
\usepackage{xcolor}
\usepackage{wasysym}
\usepackage{pgfplotstable}
\usepackage[normalem]{ulem}
\usetikzlibrary{arrows.meta}
\usepackage{smartdiagram}

\usepackage{enumitem}
\usepackage{amsfonts}
\usepackage{enumitem,kantlipsum}
\usepackage{textcomp}
\usepackage{stfloats}
\usepackage{url}
\usepackage{caption}
\usepackage{graphics}
\usepackage{tabularx}
\usepackage[dvipsnames,table]{xcolor}
\usepackage{verbatim}
\usepackage{pifont}
\usepackage{float}
\usepackage{amsmath}
\usepackage{cancel}
\usepackage{libertinust1math}
\usepackage{pgfplots}
\usepackage{amsmath}
\usepackage{forest}
\usepackage{graphicx}
\usepackage{cite}
\usepackage{etoolbox}
\usepackage{graphicx}
\usepackage{float}
\usepackage{lscape}
\usepackage{longtable}
\usepackage{ltxtable}
\usepackage{hhline}
\usepackage{caption}
\usepackage{wasysym}
\usepackage{subcaption}
\usepackage{algorithm} 
\usepackage{color} 
\usepackage{colortbl}
\usepackage{algpseudocode}
\usepackage{verbatim}
\usepackage{lipsum}
\usetikzlibrary{automata, arrows.meta, positioning}
\usetikzlibrary{decorations.pathreplacing}
\usepackage{amsthm}
\usepackage{booktabs}
\usepackage{amsmath}
\usepackage{algorithm}
\usepackage{inputenc}
\usepackage{adjustbox}
\usepackage{url} 
\usepackage{multirow}
\usepackage{color,soul}
\usepackage{lipsum}
\usepackage{mathtools}
\usepackage{cuted}
\usepackage{xcolor}
\usepackage{wasysym}
\usepackage{pgfplotstable}
\usepackage[normalem]{ulem}
\usetikzlibrary{arrows.meta}
\usepackage{smartdiagram}
\usetikzlibrary{positioning, fit, calc, shapes, arrows,trees}
\usepackage{hhline}
\usetikzlibrary{positioning, fit, calc, shapes, arrows,trees}
\usepackage{hhline}
\definecolor{columbiablue}{rgb}{0.61, 0.87, 1.0}

\tikzset{%
 thick arrow/.style={
 -{Triangle[angle=120:1pt 1]},
 line width=0.8cm, 
 draw=blue!20
 },
 arrow label/.style={
 text=black,
 align=center
 },
 set mark/.style={
 insert path={
 node [midway, arrow label, node contents=#1]
 }
 }
}
\newcommand\deleted{\bgroup\markoverwith{\textcolor{red}{\rule[0.5ex]{2pt}{0.4pt}}}\ULon}

\usetikzlibrary{pgfplots.groupplots}

\newcommand\doublecheck{\textcolor{black}{\checkmark\kern-0em\checkmark}}
\newcommand\semidoublecheck{\textcolor{black}{\checkmark\kern-0em\bcancel{\checkmark}}}

\definecolor{lemon}{rgb}{1.0, 1.0, 0.13}

\newcommand\low{\cellcolor{green!60}L}
\newcommand\med{\cellcolor{lemon!80}M}
\newcommand\high{\cellcolor{red!80}H}

\begin{document}
\definecolor{BulletsColor}{rgb}{0, 0, 0.9}
\newlist{myBullets}{itemize}{1}

\setlist[myBullets]{
 label={\textbullet},
 leftmargin=*,
 topsep=0ex,
 partopsep=0ex,
 parsep=0ex,
 itemsep=0ex,
}

\newlist{myBullets1}{itemize}{1}

\setlist[myBullets1]{
 label={\textbullet},
 leftmargin=*,
 topsep=0ex,
 partopsep=0ex,
 parsep=0ex,
 itemsep=0ex,
}
\title{Cybersecurity in the Quantum Era: Assessing the Impact of Quantum Computing on Infrastructure} 

\author{Yaser Baseri, Vikas Chouhan, Ali Ghorbani
 
\IEEEcompsocitemizethanks{\IEEEcompsocthanksitem Yaser Baseri is with University of Montreal, Canada.
Email: yaser.baseri@umontreal.ca. \protect  Vikas Chouhan, and Ali Ghorbani are with Canadian Institute for Cybersecurity (CIC), University of New Brunswick, Canada.
Emails: vikas.chouhan@unb.ca; ghorbani@unb.ca. \protect} }

\maketitle
\begin{abstract}

The advent of quantum computing marks a transformative era in cybersecurity, challenging traditional cryptographic frameworks and broadening the horizons of computational capabilities. This paper navigates the transition to a quantum-resistant security framework, offering a critical analysis of encryption methods essential for the protection of critical infrastructure and cloud services in the quantum era. It meticulously evaluates the effects of quantum computing across various infrastructure layers, including applications, data, runtime, middleware, operating systems, virtualization, hardware, storage, and networks. Through a nuanced exploration of potential vulnerabilities and the evolving threat landscape conferred by quantum technologies, our study delivers strategic insights for the development of effective countermeasures. Our comprehensive inquiry underscores the pivotal shift necessitated by the disruptive potential of quantum computing, advocating for preemptive and inventive security strategies. It calls for concerted, multi-sectoral collaboration to cultivate resilient, quantum-resistant cryptographic practices. Through meticulous research, we articulate the emergent threat vectors posed by quantum technologies and evaluate post-quantum cryptographic solutions, laying the groundwork for safeguarding both current and prospective infrastructural and cloud ecosystems. Significantly, this study introduces a tailored security blueprint encompassing nine critical infrastructure components, reinforcing each domain's defenses against quantum-induced cyber threats. Our strategic vulnerability and risk assessment provide essential countermeasures, equipping stakeholders with the acumen to navigate the complex, quantum-threatened landscape. These contributions are integral for those tasked with critical infrastructure stewardship, influencing design, implementation, and policy formulation. In essence, this paper not only forecasts quantum threats but also offers a sophisticated, actionable framework for strengthening infrastructure and cloud environments against the multifaceted challenges of the quantum era.
\end{abstract}


\section{Introduction}

The emergence of quantum computing represents a paradigm shift in the landscape of infrastructure security. Quantum computing's significant implications infiltrate every layer of our digital infrastructure, casting a shadow of uncertainty over the realm of cybersecurity. Renowned for their unparalleled computational capabilities, quantum computers present a formidable challenge to the traditional cryptographic methods that have long served as the cornerstone of data protection. Notably, encryption techniques like RSA and ECC, which have historically safeguarded data integrity across myriad infrastructure and cloud systems, now face an unprecedented and imminent threat~\cite{shor1999polynomial, bernstein2009introduction, PNNLQuantumPKC}.

Quantum computing's extraordinary capabilities in tackling complex problems, such as factorizing large numbers and computing discrete logarithms, poses a significant adversary to the security of our digital infrastructure and cloud-based systems~\cite{Forbes2023,Belfer2023}. As quantum computers continue to advance in power and capacity, they assume the role of potential adversaries capable of undermining well-established encryption techniques. The repercussions of such a scenario are profound, with malicious actors potentially gaining unauthorized access to and control over critical data. This vulnerability extends its impact from individuals and organizations to entire nations, with far-reaching consequences~\cite{mosca2018cybersecurity, gidney2019factor}.

In light of this significant paradigm shift, the transition to a quantum-safe framework necessitates a comprehensive exploration of the cryptographic techniques that underpin infrastructure security. Our investigation delves deep into the complexities of quantum threats across a spectrum of infrastructure elements, encompassing applications, data, runtime, middleware, operating systems, virtualization, hardware, storage, and networks.

This study provides an in-depth exploration of the threats associated with migrating from a non-quantum-safe cryptographic state to one resilient to quantum attacks. Before organizations migrate their cryptographic infrastructure to quantum-safe algorithms, they face a range of vulnerabilities that quantum computers can exploit. These vulnerabilities include cryptographic attacks that could potentially break commonly used cryptographic algorithms such as RSA, Diffie-Hellman, or elliptic curve cryptography. Quantum computers' capabilities in tackling these algorithms could lead to data interception and decryption, exposing sensitive information like passwords or financial transactions. Identity theft becomes a significant concern as quantum computers may be used to crack digital signatures, enabling attackers to impersonate legitimate users or entities, potentially gaining unauthorized access to sensitive systems and data. Moreover, financial fraud could be perpetrated as quantum computers might compromise the cryptographic algorithms protecting financial transactions, thereby allowing attackers to steal funds, manipulate financial data, or transfer money to their accounts. Quantum computing's data manipulation capabilities raise concerns about potential alterations to critical records, financial data, or other sensitive information. Lastly, the cyber espionage landscape could see nations or organizations with access to quantum computing employing it for data theft, targeting sensitive information such as trade secrets or classified data~\cite{NISTPQC, NatureQuantumSecrets, CISAPQC, TechTargetQuantum, ISACAQuantumThreat}.

Even after migrating to quantum-safe algorithms, organizations may still be vulnerable to specific types of attacks that do not rely on breaking encryption. These threats include denial-of-service attacks that could be launched using quantum computers to disrupt critical infrastructure, potentially causing disruptions in power grids or financial systems. Cryptographic protocol attacks become a concern, with attackers targeting the implementation of quantum-safe algorithms to exploit weaknesses and gain access to sensitive information. Social engineering attacks, such as phishing, continue to pose risks as they rely on user deception to trick individuals into disclosing sensitive information. Furthermore, quantum computers could be used to create more advanced and stealthy malware that is challenging to detect, posing a significant risk to post-migration infrastructure~\cite{NatureQuantumSecrets, CISAPQC, TechTargetQuantum}.

It is crucial to note that while quantum computing poses a long-term threat to encryption, practical quantum attacks are not yet widespread, and the timeline for when they may become a significant concern remains uncertain. Organizations are actively researching and developing quantum-resistant encryption methods and security protocols to mitigate these potential risks~\cite{bernstein2017postquantum}.

This paper embarks on an exhaustive examination of the cyber impact of quantum computing on infrastructure, emphasizing the vulnerabilities arising from quantum threats. We employ established criteria and STRIDE (Spoofing, Tampering, Repudiation, Information Disclosure, Denial of Service, and Elevation of Privilege) mapping to identify, evaluate, and prioritize potential threats to critical assets, encompassing information, technology, and physical infrastructure~\cite{khan2017stride,khalil2023threat}. Simultaneously, we underscore the vital importance of quantum-resistant measures to shield against impending attacks.
In summary, the advent of quantum computing is reshaping the cybersecurity landscape, introducing new challenges to traditional cryptographic methods and pushing the boundaries of computational capabilities. Our study systematically identifies and assesses vulnerabilities and threats both before and after migration to quantum-safe algorithms, providing valuable insights for the development of appropriate countermeasures. The findings of this study significantly advance our understanding of the impact of quantum computing on infrastructure, offering practical guidance for those engaged in the design, implementation, and policy formulation related to critical infrastructure. This comprehensive study marks a pivotal stride toward enhancing the security of networked environments in the era of quantum computing. 

\begin{figure}[!htbp]
\centering
{\includegraphics[width=0.8\linewidth, trim=35 380 140 60,clip]{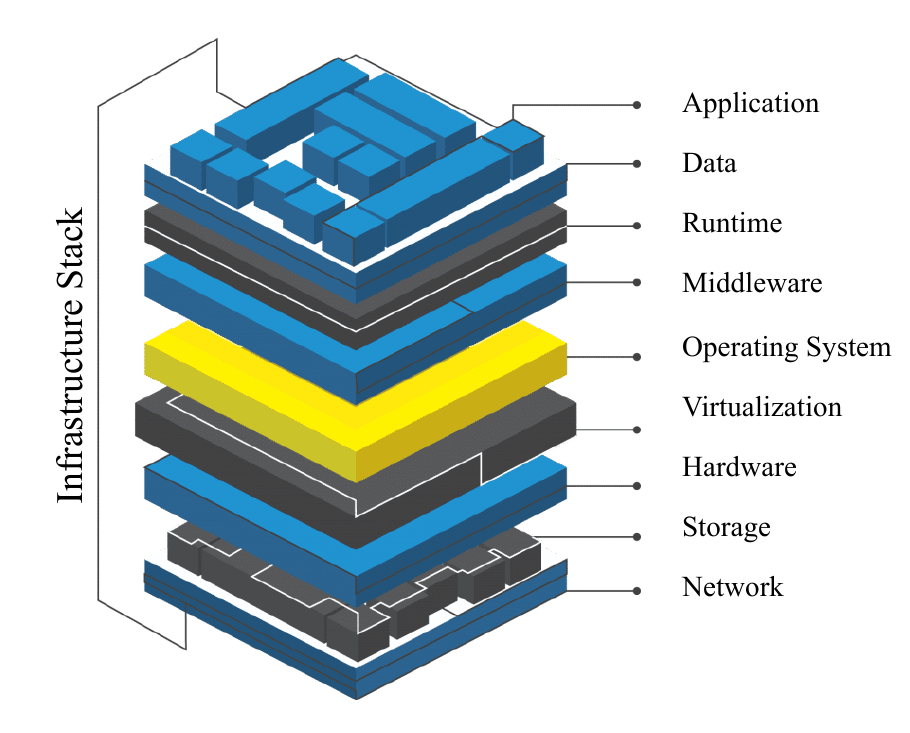}}
\caption{Infrastructure  Stack Components}
 \label{fig:stack}
\end{figure}

\subsection{Motivation}
Building upon the insights presented in the introduction, this research is motivated by the imperative to address the profound cybersecurity challenges brought forth by the rapid advancements in quantum computing, especially within critical infrastructure and cloud services. Quantum computing, with its potential for exponentially greater computational power, represents a formidable threat to traditional cryptographic methods, which are foundational to the security of data in terms of its integrity, confidentiality, and availability.

Quantum adversaries, leveraging these advanced capabilities, could feasibly exploit vulnerabilities in systems that are not quantum-safe, leading to severe consequences like data breaches, identity theft, and unauthorized access. The implications are particularly critical for cloud services, handled by industry giants like Amazon, Google, and Microsoft, where the security of massive volumes of sensitive data is at stake.

Despite the uncertainty surrounding the timeline of practical quantum attacks, this research proactively delves into identifying vulnerabilities and crafting potential countermeasures. We aim to thoroughly comprehend and anticipate the multifaceted impact of quantum computing various infrastructure layers, as illustrated in Figure~\ref{fig:stack}, as well as on cloud services. This study not only highlights potential threats but also explores the readiness of cloud providers and the implementation of quantum-resistant strategies, offering practical guidance to reinforce infrastructure security. 
By doing so, our research not only addresses an urgent need in the face of emerging quantum threats but also contributes vital insights to the broader discourse on protecting our interconnected digital world. This research, therefore, stands as a crucial step in preparing and fortifying digital infrastructure against the impending era of quantum computing.

\subsection{Contribution}
This paper significantly contributes to the domain of infrastructure security in the era of quantum computing, focusing on the development and understanding of quantum-resistant security measures. The key contributions include:

\begin{itemize}
\item \textbf{In-Depth Threat Analysis across Infrastructure Layers:} 
Our exhaustive examination provides a multi-layered analysis of the potential security threats emerging from quantum computing. This analysis extends across various infrastructure stages, covering both pre- and post-quantum cryptography migration scenarios. It presents a nuanced understanding of the evolving threat landscape, identifying specific vulnerabilities and attack vectors that quantum adversaries may exploit.

\item \textbf{Application of STRIDE for Quantum Threat Modeling:} 
We enhance our threat analysis by incorporating the STRIDE model, offering a structured approach to identify and contextualize threats in a quantum computing context. This methodology is particularly effective during the crucial transition period where systems operate under both classical and quantum-safe states. It enables a deeper comprehension of the quantum-specific threats and assists in formulating strategic responses, ensuring uninterrupted business operations during these transitional phases.

\item \textbf{Customized Risk Assessment Framework for Infrastructure Migration:} 
Our proposed risk assessment framework is custom-designed for evaluating the specific risks associated with migrating to a quantum-resistant infrastructure. Building upon our detailed threat analysis, the framework offers criteria to systematically assess, interpret, and mitigate the risks related to adopting quantum-safe algorithms and systems. It serves as a strategic tool for decision-makers, guiding risk management in a quantum-threatened landscape.
\end{itemize}

These contributions collectively mark a significant advancement in securing infrastructures against quantum computing threats. They provide vital insights and methodologies for stakeholders, shaping the development of quantum-resistant security measures, guiding infrastructure adaptation, and influencing policy decisions. Therefore, this paper plays a critical role in fortifying the security posture of networked systems in anticipation of quantum computing advancements.

\subsection{Organization}
This paper is organized as follows: Section~\ref{sec:related-works} provides a background on the intersection of quantum computing and cybersecurity, with a focus on quantum-resistant cryptographic methods. Section~\ref{sec:risk} evaluates the cybersecurity risks associated with quantum computing, encompassing both classical and emerging post-quantum cryptography, and strategies for developing quantum-safe infrastructures. Section~\ref{sec:pre} discusses the pre-migration phase, examining the impact of quantum computing on cybersecurity in critical infrastructures like cloud computing and the challenges it poses to traditional cryptography. Section~\ref{sec:post} delves into the post-migration phase, analyzing how digital security is adapting through quantum-resistant cryptography and its implications for key infrastructures, particularly cloud environments. The paper concludes in Section~\ref{sec:conclusion} with a summary and future research directions, offering consolidated insights into the role of quantum computing in cybersecurity.

\section{Related Works}\label{sec:related-works}
This section comprehensively reviews the advancements in quantum computing, the evolution of post-quantum cryptography (PQC), and their cumulative impact on digital infrastructure and cybersecurity. It emphasizes the transition of quantum computing from theoretical concepts to applications that directly challenge traditional cryptographic protocols. The section also highlights gaps in current research, particularly in addressing the range and complexity of quantum-induced cyber threats to both present and future infrastructure, including cloud ecosystems.

\subsection{Recent Trends and Future Directions in Quantum-Resistant Security}

Recent progress in quantum computing raise concerns about its potential threat to conventional cryptographic systems. Mashatan~\cite{mashatan2022} emphasizes the critical need for a quantum-resistant roadmap in cybersecurity, including the anticipation of timelines for certifying quantum-resistant standards and preparation against potential quantum attacks.
Barker et al.~\cite{barker2021migration} propose a comprehensive framework emphasizing a systematic shift to post-quantum cryptography, highlighting the need for robust testing, well-planned integration timelines, and a thorough strategy for achieving quantum-safe enterprise systems in response to the challenges posed by quantum computing. 
Akter et al.~\cite{shapna2023quantum} provide a comprehensive survey of Quantum Cryptography for Enhanced Network Security, shedding light on the research landscape and future directions in this domain.
These studies collectively represent a concerted effort to navigate the evolving cybersecurity landscape in the quantum era, aiming to pave the way for robust solutions in quantum-resistant security.

\subsection{Evolution of Quantum Computing and Cryptography}
Quantum computing has swiftly transitioned from theoretical exploration to practical applications, with significant implications in the field of cryptography. The groundbreaking work by Shor~\cite{shor1994algorithms} revealed the vulnerability of conventional cryptographic protocols, such as RSA and ECC, against quantum computing attacks, specifically how quantum algorithms could exploit mathematical shortcuts for breaking these systems. This breakthrough has catalyzed the development of post-quantum cryptography (PQC), which is focused on designing algorithms secure against both classical and quantum computational threats. Recent advancements in quantum computing technology, such as those by Google and IBM, illustrate an accelerated approach towards practical applications, underscoring the urgency of PQC development~\cite{arute2019quantum, castelvecchi2017quantum, bernstein2017postquantum}.

\subsection{Development and Standardization of Post-Quantum Cryptography}
In response to quantum threats, research and standardization in PQC have accelerated. Organizations like the National Institute of Standards and Technology (NIST) are evaluating a variety of cryptographic methods, with lattice-base, hash-based, code-based, and isogeny-based cryptography each offering unique strengths and potential use cases. For instance, lattice-based cryptography is gaining attention due to its efficiency and resistance to quantum attacks. NIST's ongoing efforts to standardize quantum-resistant algorithms are shaping the future landscape of secure digital communication~\cite{NIST_2022July, mosca2018cybersecurity, nist2023postquantum, nist2022selectedPQalgo}. The contributions by Kumar and Garhwal~\cite{kumar2021state} in exploring new algorithmic paradigms have further enriched this domain, including developments in algorithm agility and adaptability.

\subsection{Impact of Quantum Computing on Infrastructure Security}

The rapid evolution of quantum computing poses significant challenges to digital infrastructure, necessitating a comprehensive understanding of potential vulnerabilities introduced by quantum technologies. Recent advancements in quantum computing technology by Google and IBM underscore the urgency of addressing quantum threats. Lindsay~\cite{lindsay2020demystifying} analyzes quantum computing's potential impact on cryptographic protocols, providing insights into the strategic alignment necessary to fortify security in the quantum era. Mangla et al.~\cite{mangla2023mitigating} focus on vulnerabilities in 5G networks and anticipated security challenges for future 6G networks, providing insights into fortifying security amid the evolving telecommunications landscape.

\subsection{Mitigating Quantum Threats in Digital Infrastructures}

Addressing the multifaceted challenges posed by quantum advancements, pivotal works offer diverse perspectives. Lindsay's analysis~\cite{lindsay2020demystifying} emphasizes the intricate interplay between technological infrastructure and organizational institutions. Mangla et al.~\cite{mangla2023mitigating} focus on vulnerabilities in 5G networks.  Mosca and  Piani's Quantum Threat Timeline Report 2022 is expected to provide a chronological assessment of the quantum threat landscape~\cite{mosca2022quantum}. Faruk et al.~\cite{faruk2022review} explore quantum computing's dual nature as both a potential threat and a solution in the cybersecurity domain.



\subsection{Strategies of Major Cloud Providers in Securing infrastructure Against Quantum Threats}
Quantum computing presents groundbreaking opportunities and significant cybersecurity challenges, impacting cloud infrastructure. Major cloud providers such as Amazon Web Services (AWS)~\cite{aws2023quantum}, Microsoft Azure~\cite{azure2023quantum}, and Google Cloud Platform (GCP)~\cite{google2023quantum}, are actively developing strategies to address these challenges. 
These providers focus on quantum-resistant cryptography (PQC)~\cite{bernstein2017postquantum} to secure against quantum computing threats. They adopt hybrid cryptographic solutions~\cite{driscoll-pqt-hybrid-terminology-02}, combining current encryption methods with quantum-resistant algorithms~\cite{mosca2018cybersecurity}, and emphasize cryptographic agility~\cite{rossenberg2019postquantum}.

Providers are balancing the trade-offs between security and computational performance, as some quantum-resistant algorithms can be less efficient~\cite{mosca2018cybersecurity}. Transitioning to PQC involves substantial system updates, demanding careful integration to maintain system integrity and backward compatibility~\cite{CFDIR}.
Their efforts encompass all layers of infrastructure: developing quantum-safe applications~\cite{chen2016report}, protecting data with quantum-resistant encryption, securing runtime environments~\cite{mattsson2021quantum}, updating middleware protocols, integrating quantum-safe features in operating systems, fortifying virtualization layers~\cite{techtarget2021cloud}, implementing quantum-resistant hardware security modules~\cite{crypto4aHSM}, enhancing storage encryption~\cite{entrust2023encryption}, and adopting quantum-safe networking protocols~\cite{ibmQuantumSafe}.
These strategies ensure a comprehensive examination of quantum threats across the entire stack, with a focus on collaboration with hardware vendors, cryptographic communities, and researchers~\cite{gyongyosi2019survey}. Compliance with global quantum-safe standards, like those spearheaded by NIST~\cite{nist2023quantum}, is also emphasized. Providers update their threat models to encompass potential quantum computer attacks, reinforcing their security posture~\cite{gidney2021quantum}.

In conclusion, AWS, Azure, and GCP are addressing the challenges posed by quantum computing~\cite{gyongyosi2019survey}. Their strategies focus on implementing quantum-resistant technologies, collaborative research, standardization, and proactive security measures, highlighting their commitment to safeguarding infrastructure in the quantum era. However, the evolving nature of quantum technologies means that these strategies will require ongoing updates and vigilance.

\subsection{Gaps in Current Research and Our Focus}
While there has been significant progress in PQC and understanding the broader implications of quantum computing, there remains a critical gap in detailed analyses of quantum-induced cyber threats, particularly those affecting existing and future infrastructures and cloud environments. Previous research has mostly centered on general aspects of PQC and quantum computing's implications, often missing the finer details of these emerging challenges. Our research aims to fill this gap, delving into the nuanced threats posed by quantum advancements and formulating novel strategies to protect infrastructures against them. This work ventures into a vital yet scarcely explored domain, fortifying defenses in the rapidly evolving quantum computing and cybersecurity landscape.

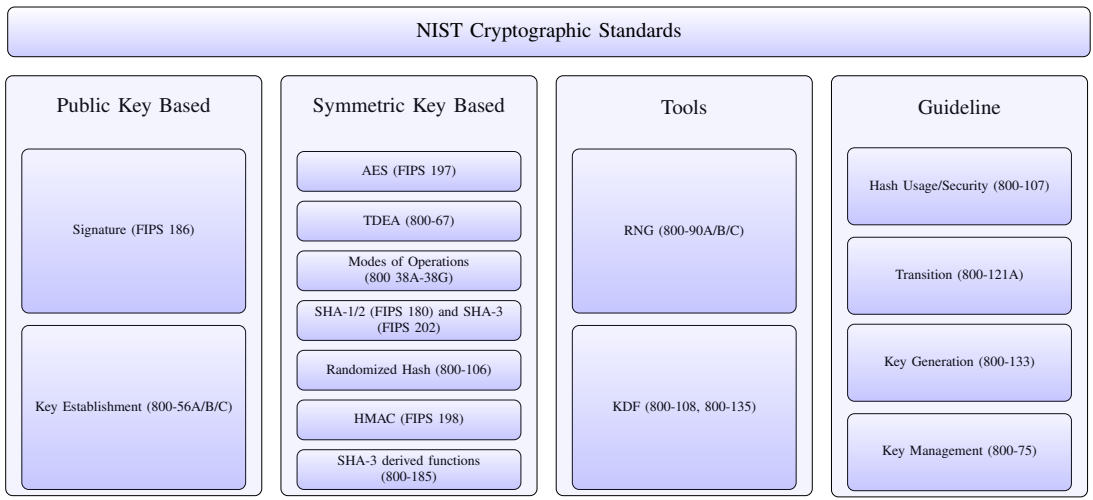
\begin{figure*}[!ht]
 \begin{center} 
\resizebox{0.8\linewidth}{!}{
\begin{tikzpicture}[
 title/.style={minimum height=1cm,minimum width=4.9cm,font = {\large}},
 body/.style={draw,top color=white, bottom color=blue!20, rounded corners,minimum width=4.5cm,,minimum height=0.8cm,font = {\footnotesize}},
 typetag/.style={rectangle, draw=black!100, anchor=west}]
 \node (d0) [draw,top color=white, bottom color=blue!20, rounded corners,minimum height=1cm,minimum width=1.2\textwidth,font = {\large}] at (0,0) {NIST Cryptographic Standards
};
 
\node (d3) [title,below=of d0.center,xshift=0.15\textwidth] {Tools};
 \node (d31) [body,below=of d3.west, typetag, xshift=2mm,yshift=-1.5cm,minimum height=3.3cm] {\begin{minipage}[c]{4.00cm}\centering
RNG (800-90A/B/C)
\end{minipage}};
 \node (d32) [body,below=of d31.west, typetag,yshift=-2.55cm,minimum height=3.3cm] {\begin{minipage}[c]{4.00cm}\centering
KDF (800-108, 800-135)
\end{minipage}};
\node [top color=blue!40, bottom color=blue!40, rounded corners,minimum height=1cm,draw=black!100,fill opacity=0.1, fit={(d3) (d31) (d32)}] {};
 
\node (d2) [title, left of=d3,xshift=-0.25\textwidth] {Symmetric Key Based
};
 \node (d21) [body,below=of d2.west, typetag, yshift=-0.3cm, xshift=2mm] {\begin{minipage}[c]{4.00cm}\centering
AES (FIPS 197)
\end{minipage}};
 \node (d22) [body,below=of d21.west, typetag] {\begin{minipage}[c]{4.00cm}\centering
TDEA (800-67)
\end{minipage}};
 \node (d23) [body,below=of d22.west, typetag] {\begin{minipage}[c]{4.00cm}\centering
Modes of Operations\\ (800 38A-38G)
\end{minipage}};
\node (d24) [body,below=of d23.west, typetag] {\begin{minipage}[c]{4.00cm}\centering
SHA-1/2 (FIPS 180) and SHA-3 (FIPS 202) 
\end{minipage}};
\node (d25) [body,below=of d24.west, typetag] {\begin{minipage}[c]{4.00cm}\centering
Randomized Hash (800-106)
\end{minipage}};
\node (d26) [body,below=of d25.west, typetag]{\begin{minipage}[c]{4.00cm}\centering
HMAC (FIPS 198)
\end{minipage}};
\node (d27) [body,below=of d26.west, typetag]{\begin{minipage}[c]{4.00cm}\centering
SHA-3 derived functions (800-185)
\end{minipage}};
\node [top color=blue!40, bottom color=blue!40, rounded corners,minimum height=1cm,draw=black!100,fill opacity=0.1, fit={(d2) (d21) (d22) (d23) (d24) (d25) (d26) (d27)}] {};

\node (d1) [title,left of=d2,xshift=-0.25\textwidth] {Public Key Based
};
 \node (d11) [body,below=of d1.west, typetag, xshift=2mm, yshift=-1.5cm,minimum height=3.3cm] {\begin{minipage}[c]{4.00cm}\centering
Signature (FIPS 186)
\end{minipage}};
 \node (d12) [body,below=of d11.west, typetag,yshift=-2.55cm,minimum height=3.3cm] {
\begin{minipage}[c]{4.00cm}\centering
Key Establishment (800-56A/B/C)
\end{minipage}};

\node [top color=blue!40, bottom color=blue!40, rounded corners,minimum height=1cm,draw=black!100,fill opacity=0.1, fit={(d1) (d11) (d12)}] {}; 
 
\node (d4) [title, right of=d3,xshift=0.25\textwidth] {Guideline};
 \node (d41) [body,below=of d4.west, typetag, yshift=-0.6cm, xshift=2mm,minimum height=1.55cm] {\begin{minipage}[c]{4.00cm}\centering
Hash Usage/Security (800-107)
\end{minipage}};
 \node (d42) [body,below=of d41.west, typetag, yshift=-0.8cm,minimum height=1.55cm] {\begin{minipage}[c]{4.00cm}\centering
Transition (800-121A)
\end{minipage}};
 \node (d43) [body,below=of d42.west, typetag, yshift=-0.75cm,minimum height=1.55cm] {\begin{minipage}[c]{4.00cm}\centering
Key Generation (800-133)
\end{minipage}};
\node (d44) [body,below=of d43.west, typetag, yshift=-0.8cm,minimum height=1.55cm] {\begin{minipage}[c]{4.00cm}\centering
Key Management (800-75)
\end{minipage}};
\node [top color=blue!40, bottom color=blue!40, rounded corners,minimum height=1cm,draw=black!100,fill opacity=0.1, fit={(d4) (d41) (d42) (d43) (d44)}] {};

 \end{tikzpicture}
}
\caption{NIST Cryptographic Standards}
\label{fig:Temp-tree-standards}
\end{center}
\end{figure*}

 \section{Cryptographic Standards and Quantum Computing: Cyber Impact and Risk Assessment}\label{sec:risk}
This paper presents an elaborate security framework designed to systematically tackle the broad range of cybersecurity threats precipitated by the advent of quantum computing. These threats pertain to both current and forthcoming infrastructure, including cloud-based environments. Our framework introduces a detailed security blueprint, focusing on nine critical infrastructural elements: applications, data, runtime, middleware, operating systems, virtualization, hardware, storage, and networks. This framework aims to address the vulnerabilities and risks associated with these components in the face of emerging quantum computing capabilities, ensuring robust protection for both existing and future cloud and infrastructure systems.

Before we dive into our discussion, it is crucial to understand the seriousness of the quantum computing threat. The vulnerability of cryptographic systems is at the core of the risks to infrastructure and cloud security. Therefore, we need to assess how quantum computing impacts both traditional and post-quantum cryptography, which is considered a potential solution for quantum-safe cryptographic systems. Evaluating these risks is essential for companies to effectively prioritize threats by considering the potential consequences and the likelihood of vulnerabilities being exploited. In the context of transitioning to quantum-safe solutions, we use a risk assessment methodology guided by the recommendations of the National Institute of Standards and Technology (NIST).

Our security framework is specifically designed to counter the emerging threats introduced by quantum computing in the field of current and future cryptography. 
In Figure~\ref{fig:Temp-tree-standards}, you can see an overview of selected cryptographic standards provided by NIST~\cite{nist_sp_800_57, nist_sp_800_38, nist_sp_800_90, nist_sp_800_131A, fips_140_2, fips_140_3, nist_sp_800_56}. 
With the impending arrival of quantum computers with significant capabilities, their impact on both public and symmetric cryptographic systems is unavoidable. Even the new quantum-safe cryptographic methods being considered for standardization by NIST are not entirely immune to vulnerabilities. In this section, we assess the security risks associated with the evolving threats posed by quantum computing to existing and potential cryptographic solutions.

\subsection{Classic Cryptographic Standards and Quantum Computing: Assessing Cyber Risks}

Cryptographic standards, the cornerstone of security in diverse applications and communications, primarily employ classical algorithms, both asymmetric and symmetric. These algorithms face a growing threat from quantum computing, which could break or weaken them significantly. The hard problems underpinning modern asymmetric cryptography, the cornerstone of secure communication, may no longer pose a substantial challenge for quantum attackers equipped with sufficiently powerful quantum computers. Utilizing established quantum algorithms, such as Shor's algorithm, these attackers can potentially breach the security of widely used asymmetric cryptographic systems. 
Similarly, symmetric cryptographic systems' security is also at risk. Quantum algorithms like Grover's algorithm and the Brassard et al.'s algorithm can undermine the security of symmetric cryptographic algorithms and communication protocols. Consequently, this section delves into the vulnerabilities and strengths of classical cryptographic algorithms as we approach the need for migration to potentially quantum-safe cryptographic alternatives.\\

\begin{enumerate}[wide, font=\itshape, labelwidth=!, labelindent=0pt, label*=\textit{A}.\arabic*.]
\item \textit{Quantifying Quantum Threats:}
 To understand the risks associated with quantum migration, it is imperative to predict the emergence of quantum computers and the resultant risks to classical cryptosystems. 
Our analysis examines the timeline for quantum computers to appear within the next 5 to 30 years. This analysis is built on a cumulative likelihood of significant quantum threats to classical cryptosystems.

\begin{figure}[!htbp]
\centering
\frame{\includegraphics[trim=0.2cm 0.1cm 0.2cm 0.2cm, clip=true, width=\linewidth, keepaspectratio]{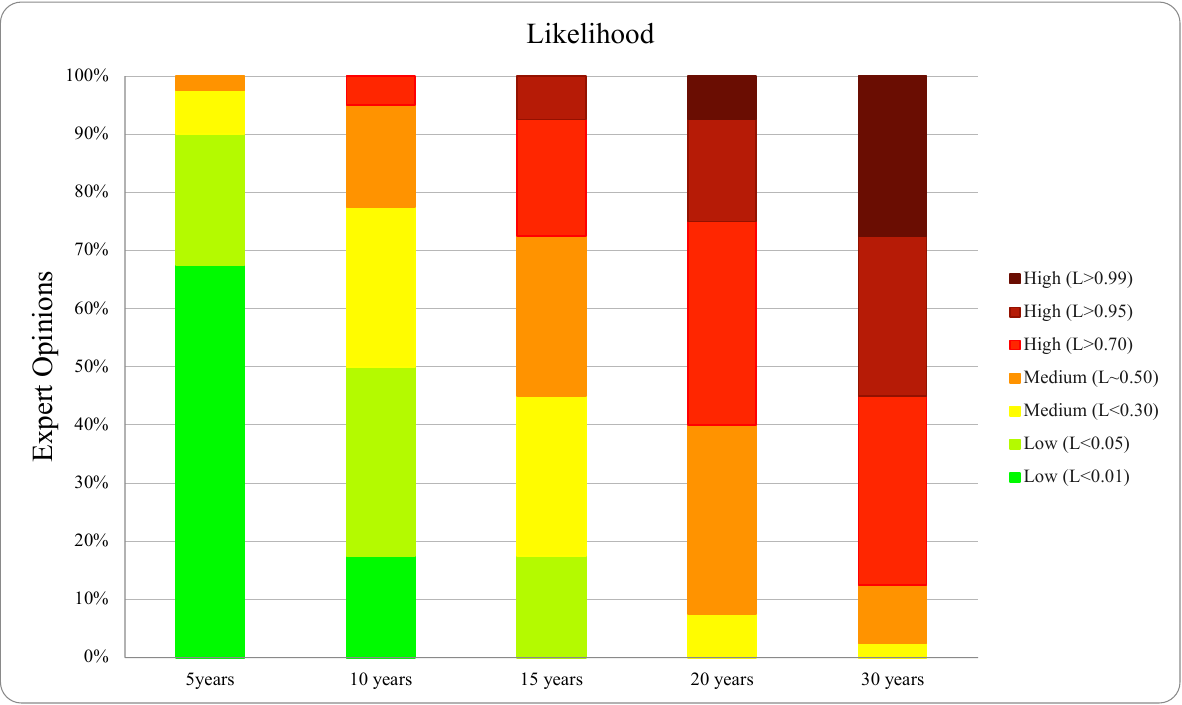}}
\caption{Cumulative Expert Opinions Related to Quantum Threat to Classic Cryptography}
\label{fig:chart1}
\end{figure}

Figure~\ref{fig:chart1} summarizes this evolution, incorporating insights from multiple quantum experts regarding the quantum threat timeline~\cite{mosca2022quantum}. 
The ``quantum threat" is defined as the probability of breaking RSA-2048 within 24 hours using a quantum machine. These assessments can be extended to evaluate the likelihood of breaking other cryptographic algorithms based on their quantum security level, as presented in Figure~\ref{fig:chart2}.\\

 \item \textit{Expected Likelihood of Quantum Threat:} To assess the ``expected likelihood of the quantum threat for classical cryptosystems" over various periods (5, 10, 15, 20, and 30 years), we aggregate predictions made by different experts who participated in the poll~\cite{mosca2022quantum}. For each period, such as 5 years, we calculate the expected likelihood by multiplying the agreed-upon likelihoods of predictions for that period by the probability of those predictions, then summing them up.
\begin{equation*}
    E_{\text{period}_j}[\text{likelihood}] = \sum_{\omega_i\subseteq [0,1]} \text{likelihood}_{\text{period}_j} (\omega_i) \times \text{Pr}_{\text{period}_j} (\omega_i)
\end{equation*}
Here, $\omega_i$ represents subsets of $[0,1]$, with their union equaling $[0,1]$. $\text{Pr}_{\text{period}_j} (\omega_i)$ for each period $\text{period}_j$ is determined by the fraction of expert opinions that agreed on prediction $\omega_i$ for that period, relative to the total number of predictions.

\begin{table*}[!hbpt]
\caption{Classic Cryptographic Standards and Quantum Computing: Assessing Cyber Risks}
\small
\label{tab:Pre-Migration-Alg}
\resizebox{\textwidth}{!}{%
\begin{tabular}{|l|l|l|l|l|l|p{0.3\linewidth}|l|l|l|p{0.35\linewidth}|}
\hline
\multirow{2}{*}{\textbf{$ $Crypto Type$ $}} & \multirow{2}{*}{\textbf{$ $Algorithms$ $}} & \multirow{2}{*}{\textbf{$ $Variants$ $}} & \multirow{2}{*}{\textbf{$ $Key Length (bits)$ $}} & \multirow{2}{*}{\textbf{$ $Classic Strength (bits)$ $}} & \multirow{2}{*}{\textbf{$ $Quantum Strength (bits)$ $}} & \multirow{2}{*}{\textbf{$ $Vulnerabilities$ $}} & \multirow{2}{*}{\textbf{L}} & \multirow{2}{*}{\textbf{I}} & \multirow{2}{*}{\textbf{R}} & \multirow{2}{*}{\textbf{$ $Recommended QC-Resistant Solutions$ $}} \\ 
 & & & & & & & & & & \\ \hline
\multirow{9}{*}{Asymetric} & \multirow{3}{*}{ECC~\cite{turner2010use}} & ECC 256 & 256 & \multicolumn{1}{l|}{128} & 0 & \multirow{3}{*}{{\begin{minipage}{\linewidth}
Broken by Shor's Algorithm~\cite{shor1994algorithms}.
\end{minipage}}} & \med &\high&\high& \multirow{8}{*}{{\begin{minipage}{\linewidth}Algorithms presented in Table~\ref{tab:side}.\end{minipage}}}\\\cline{3-6}\cline{9-10}
 & & ECC 384 & 384 & \multicolumn{1}{l|}{256} & 0 & & \med& \high &\high& \\\cline{3-6}\cline{9-10}
 & & ECC 521 & 521 & \multicolumn{1}{l|}{256} & 0 & &\med & \high&\high& \\ \cline{2-10}
 & \multirow{2}{*}{FFDHE~\cite{gillmor2016negotiated}} & DHE2048 & 2048 & \multicolumn{1}{l|}{112} & 0 & \multirow{2}{*}{{\begin{minipage}{\linewidth}
Broken by Shor's Algorithm~\cite{shor1994algorithms}.
\end{minipage}}} &\med & \high&\high &\\\cline{3-6}\cline{9-10}
 & & DHE3072 & 3072 & \multicolumn{1}{l|}{128} & 0 & & \med& \high& \high&\\ \cline{2-10}
 & \multirow{3}{*}{RSA~\cite{moriarty2016pkcs}} & RSA 1024 & 1024 & \multicolumn{1}{l|}{80} & 0 & 
 \multirow{3}{*}{{\begin{minipage}{\linewidth}
Broken by Shor's Algorithm~\cite{shor1994algorithms}.
\end{minipage}}}
& \med& \high& \high & \\\cline{3-6}\cline{9-10}
 & & RSA 2048 & 2048 & \multicolumn{1}{l|}{112} & 0 & & \med& \high &\high& \\\cline{3-6}\cline{9-10}
 & & RSA 3072 & 3072 & \multicolumn{1}{l|}{128} & 0 & & \med& \high &\high& \\ \hline
\multirow{4}{*}{Symmetric} & \multirow{3}{*}{AES~\cite{schaad2003use}} & AES 128 & 128 & \multicolumn{1}{l|}{128} & 64 & \multirow{3}{*}{{\begin{minipage}{\linewidth}
Weakened by Grover's Algorithm~\cite{grover1996fast}.
\end{minipage}}}
& \med& \med & \med & \multirow{3}{*}{{\begin{minipage}{\linewidth}Larger key sizes are needed.\end{minipage}}} \\
\cline{3-6}\cline{9-10}
 & & AES 192 & 192 & \multicolumn{1}{l|}{192} & 96 & & \med & \med &\med & \\\cline{3-6}\cline{9-10}
 & & AES 256 & 256 & \multicolumn{1}{l|}{256} & 128 & & \med& \low&\low& \\ 
 \cline{2-11}
 & \multirow{3}{*}{SHA2~\cite{eastlake2011us}} & SHA 256 & - & \multicolumn{1}{l|}{128} & {85} & \multirow{3}{*}{{\begin{minipage}{\linewidth}
Weakened by Brassard et al.'s Algorithm~\cite{brassard1997quantum}.
\end{minipage}}} & \med& \med &\med & \multirow{6}{*}{{\begin{minipage}{\linewidth}Larger hash values are needed.\end{minipage}}} \\\cline{3-6}\cline{9-10} 
 & & SHA 384 & - & \multicolumn{1}{l|}{192} & 128 & & \med & \low&\low& \\\cline{3-6}\cline{9-10}
 & & SHA 512 & - & \multicolumn{1}{l|}{256} & 170 & & \med& \low&\low& \\ \cline{2-10}
 & \multirow{3}{*}{SHA3~\cite{eastlake2011us}} & SHA3 256 & - & \multicolumn{1}{l|}{128} & {85} & \multirow{3}{*}{{\begin{minipage}{\linewidth}
Weakened by Brassard et al.'s
Algorithm~\cite{brassard1997quantum}.
\end{minipage}}} & \med& \med &\med & \\\cline{3-6}\cline{9-10} 
 & & SHA3 384 & - & \multicolumn{1}{l|}{192} & 128 & & \med & \low&\low & \\\cline{3-6}\cline{9-10}
 & & SHA3 512 & - & \multicolumn{1}{l|}{256} & 170 & & \med& \low&\low & \\ \hline
\end{tabular}%
\vspace{-0.2cm}}
\end{table*}

Our likelihood assessment categorizes quantum threat likelihood into three levels: low, medium, and high. As depicted in Figure~\ref{fig:chart2}, the expected likelihood of a quantum threat within 10 years is low, within 15 years is medium, and beyond 20 years is high. Our evaluation adopts a medium likelihood within 15 years for the quantum threat to classical cryptosystems. This assumption can be adapted for other timeframes.\\

\begin{figure}[!htbp]
\centering
\frame{\includegraphics[trim=0.2cm 0.1cm 0.1cm 0.5cm, clip=true, width=\linewidth, keepaspectratio]{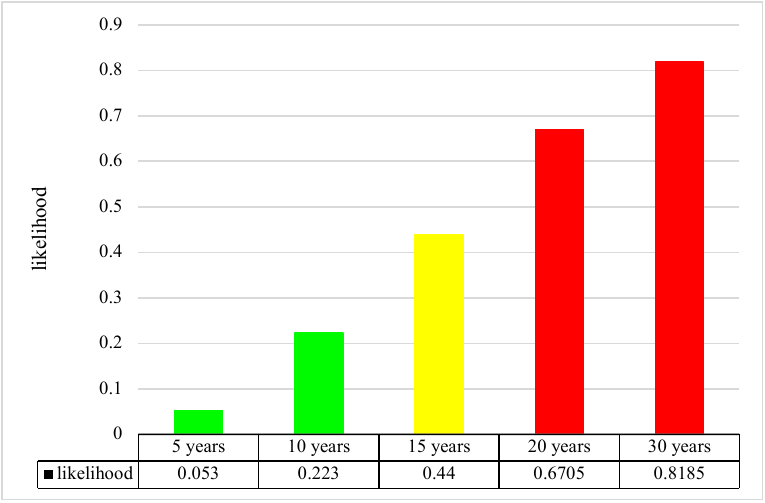}}
\caption{Expected Likelihood of Quantum Threat for Classic Cryptography Within 30 Years}
\label{fig:chart2}
\end{figure}

 \item \textit{Quantum Impact Assessment:}
To conduct a classic algorithmic level risk assessment, we evaluate the impact of quantum threats on different classic cryptographic algorithms. The impact is determined based on the quantum security strength of each classic algorithm, as illustrated in Figure~\ref{fig:classic-impact}. An impact is considered high if the algorithm's quantum strength is less than 64 bits, low if it is greater than or equal to 128 bits, and medium if it falls between these values. The final risk assessment combines both the likelihood and impact and is presented in Table~\ref{tab:Pre-Migration-Alg}.\\

\begin{figure}[!htbp]
\centering
\frame{\includegraphics[trim=0.1cm 0.1cm 0.1cm 0.55cm, clip=true, width=\linewidth, keepaspectratio]{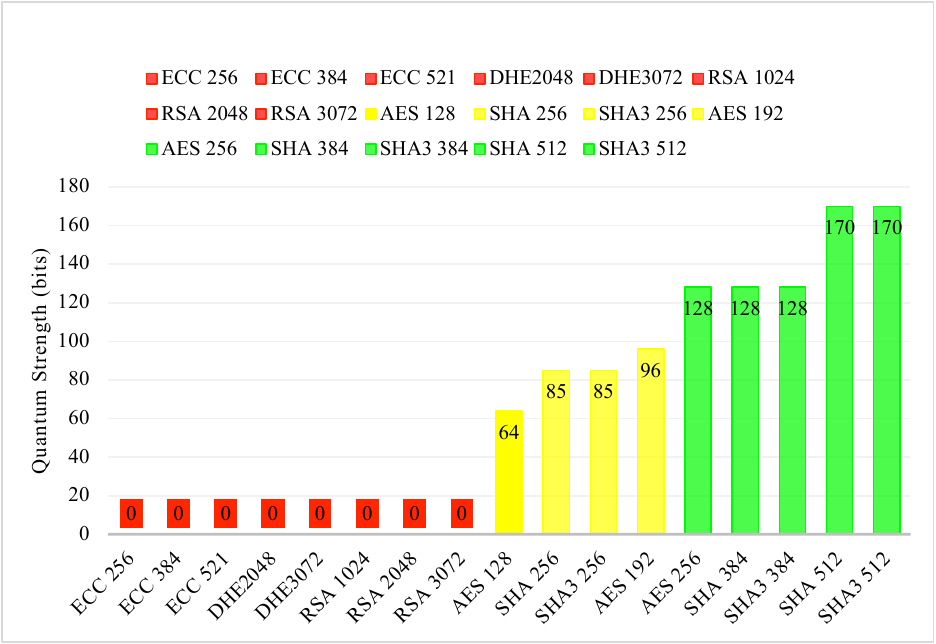}}
\caption{Expected Impact of Quantum Threat for Classic Cryptography}
\label{fig:classic-impact}
\end{figure}

 \item \textit{Risk Evaluation:}
\begin{figure}[!h]
\renewcommand{\arraystretch}{1.8}
 \small
\centering{
\resizebox{0.9\linewidth}{!}{%
\subfloat{}{\begin{math}
 \qquad \raisebox{-1\normalbaselineskip}{Likelihood$\ \left\{\rule{0pt}{3\normalbaselineskip}\right.$}
\hspace{-0.15cm}
\begin{tabular}{p{0.1cm}}\\
\multicolumn{1}{l}{Low} \\ 
\multicolumn{1}{l}{Medium} \\
\multicolumn{1}{l}{High}
\end{tabular}
\end{math}
}
\hspace{-0.3cm}
\begin{math}
 \overbrace{\
\begin{tabular}{p{1.5cm}p{1.5cm}p{1.5cm}}
Low & Medium & High \\ \hline
\multicolumn{1}{|l|}{\cellcolor{green!60}Low} & \multicolumn{1}{l|}{\cellcolor{green!60}Low} & \multicolumn{1}{l|}{\cellcolor{lemon!80}Medium} \\ \hline
\multicolumn{1}{|l|}{\cellcolor{green!60}Low} & \multicolumn{1}{l|}{\cellcolor{lemon!80}Medium} & \multicolumn{1}{l|}{\cellcolor{red!80}High} \\ \hline
\multicolumn{1}{|l|}{\cellcolor{lemon!80}Medium} & \multicolumn{1}{l|}{\cellcolor{red!80}High} & \multicolumn{1}{l|}{\cellcolor{red!80}High} \\ \hline
\end{tabular}
}^{\mbox{Impact}}
\end{math}
 }}
 \caption{Qualitative Risk Assessment based on Likelihood and Impact Levels}
 \label{fig:risk-matrix}
\end{figure}
Table~\ref{tab:Pre-Migration-Alg} offers a comprehensive insight into our findings regarding the assessment of conventional cryptographic algorithms before transitioning to quantum-safe cryptographic solutions. It provides an in-depth exploration of classical and quantum security attributes of these algorithms, their inherent vulnerabilities, and the emerging quantum threats. Furthermore, the table highlights potential quantum-resistant remedies and the attendant risks that quantum adversaries might exploit. The evaluation of these associated risks rests on a qualitative risk assessment that considers both likelihood and impact, as elucidated in Figure~\ref{fig:risk-matrix}. This holistic analysis serves as a valuable tool for assessing the security robustness of existing cryptographic algorithms and for devising effective mitigation strategies within the realm of classical cryptography.

\end{enumerate}

\subsection{Quantum-Safe Cryptographic Candidates for Standardization: Navigating Quantum Computing Risks}

\begin{table*}[!htbp]
\caption{Quantum-Safe Cryptographic Candidates for Standardization: Navigating Quantum Computing Risks}
\large
\resizebox{\textwidth}{!}{%
\begin{tabular}{|l|ll|ll|llllllll|p{1.2\textwidth}|l|l|l|}
\hline
\multirow{2}{*}{\textbf{Approaches}} & \multicolumn{2}{l|}{\textbf{Finalists}} & \multicolumn{2}{l|}{\textbf{Alternates}} & \multicolumn{8}{l|}{\textbf{Available Attacks and Countermeasures$^*$}} & \multirow{2}{*}{\textbf{Possible Countermeasures}}&\multirow{2}{*}{\textbf{L}} & \multirow{2}{*}{\textbf{I}} & \multirow{2}{*}{\textbf{R}}\\ \cline{2-13}
& \multicolumn{1}{l|}{KEM/ENC} & Signatures & \multicolumn{1}{l|}{KEM/ENC} & Signatures & \multicolumn{1}{l|}{CA} & \multicolumn{1}{l|}{TA} & \multicolumn{1}{l|}{FA} & \multicolumn{1}{l|}{SPA} & \multicolumn{1}{l|}{APA} & \multicolumn{1}{l|}{EM} & \multicolumn{1}{l|}{TMP} & CB &&&&\\ \hline
\multirow{14}{*}{Lattice-based} & \multicolumn{1}{l|}{Kyber} & & \multicolumn{1}{l|}{} & & \multicolumn{1}{l|}{} & \multicolumn{1}{l|}{} & \multicolumn{1}{l|}{\doublecheck} & \multicolumn{1}{l|}{\doublecheck} & \multicolumn{1}{l|}{\semidoublecheck} & \multicolumn{1}{l|}{\doublecheck} & \multicolumn{1}{l|}{\checkmark} & \multicolumn{1}{l|}{\doublecheck} &{\begin{minipage}{\linewidth}
\begin{myBullets}
\vspace{0.2cm}
 \item \textbf{To defend against FA:} Mask the decryption process~\cite{oder2016practical, ravi2020drop}, or check the secret and error components of the LWE instances~\cite{ravi2019number}.
 \item \textbf{To prevent SPA:} Mask the input~\cite{hamburg2021chosen}, randomize the order of executed operations~\cite{hamburg2021chosen}.
 \item \textbf{To avoid APA:} Mask the Number Theoretic Transform (NTT)~\cite{pessl2019more} (no countermeasures in place for the attacks mentioned in~\cite{dubrova2022breaking}).
 \item \textbf{To evade EM:} Mask the ECC procedures~\cite{ravi2020generic, oder2016practical}, FO transform operations~\cite{ravi2020generic}, secure the secret key~\cite{xu2021magnifying}, discard ciphertexts~\cite{xu2021magnifying}, or randomly split secrets~\cite{xu2021magnifying}.
 \item\textbf{For avoiding TMP}, no specific countermeasures mentioned for the attacks described in~\cite{ravi2021exploiting}.
 \item \textbf{To block CB:} Store the secret in the time domain instead of the frequency domain~\cite{albrecht2018cold}.\vspace{0.2cm}\end{myBullets}
\end{minipage}}&\high & \med& \high\\ \cline{2-17}
& \multicolumn{1}{l|}{} &Dilithium & \multicolumn{1}{l|}{} & & \multicolumn{1}{l|}{} & \multicolumn{1}{l|}{} & \multicolumn{1}{l|}{\doublecheck} & \multicolumn{1}{l|}{} & \multicolumn{1}{l|}{\doublecheck} & \multicolumn{1}{l|}{{\doublecheck}} & \multicolumn{1}{l|}{} & & {\begin{minipage}{\linewidth}
\begin{myBullets}
\vspace{0.2cm}
\item \textbf{To defend against FA:} 
(a) Check the secret and error components of the LWE instances~\cite{ravi2019number}, (b) Employ generic countermeasures such as double computation, verification-after-sign, and add additional randomness~\cite{bruinderink2018differential}.
\item \textbf{To avoid APA:} Implement masking techniques, including linear secret sharing~\cite{migliore2019masking} and use Boolean and arithmetic masking~\cite{marzougui2022profiling}.
\item \textbf{To evade EM:} (a) Re-order operations and embed the vulnerable addition operation within the signing procedure~\cite{ravi2019exploiting}, (b) Utilize it-slicing design for Number Theoretic Transform (NTT)~\cite{singh2022end}.

\vspace{0.2cm}\end{myBullets}
\end{minipage}}&\med & \med&\med\\ \cline{2-17} 
& \multicolumn{1}{l|}{} & Falcon & \multicolumn{1}{l|}{} & & \multicolumn{1}{l|}{} & \multicolumn{1}{l|}{\doublecheck} & \multicolumn{1}{l|}{\doublecheck} & \multicolumn{1}{l|}{\doublecheck} & \multicolumn{1}{l|}{} & \multicolumn{1}{l|}{\doublecheck} & \multicolumn{1}{l|}{} & & {\begin{minipage}{\linewidth}
\begin{myBullets}
\vspace{0.2cm}
 \item \textbf{To avoid TA:} Enhance random shuffling through the Blind-Vector algorithm~\cite{mccarthy2019bearz}, or implement sample discarding~\cite{mccarthy2019bearz}.
 \item \textbf{To defend against FA:} Duplicate computation of the signature~\cite{mccarthy2019bearz}, immediately verify after signing~\cite{mccarthy2019bearz}, and perform a zero-check on the sampled vector~\cite{mccarthy2019bearz}.
 \item \textbf{To prevent SPA:} Effectively reduce the Hamming weight gap.
 \item \textbf{To evade EM:} Conceal by maintaining a constant power consumption~\cite{karabulut2021falcon}, or mask by randomizing the intermediate values~\cite{karabulut2021falcon}.

\vspace{0.2cm}\end{myBullets}
\end{minipage}} &\med & \med&\med\\ \hline
\multirow{14}{*}{Code-based} & \multicolumn{1}{l|}{McEliece} & & \multicolumn{1}{l|}{} & & \multicolumn{1}{l|}{\doublecheck} & \multicolumn{1}{l|}{\doublecheck} & \multicolumn{1}{l|}{\semidoublecheck} & \multicolumn{1}{l|}{\semidoublecheck} & \multicolumn{1}{l|}{\doublecheck} & \multicolumn{1}{l|}{\checkmark} & \multicolumn{1}{l|}{\doublecheck} & \checkmark & {\begin{minipage}{\linewidth}\begin{myBullets}
\vspace{0.2cm}
 \item \textbf{To prevent CA:} Increase the binary code length~\cite{bernstein2008attacking} or use a decoding list to increase weight $w$~\cite{bernstein2008attacking}.
 
 \item \textbf{To defend against TA:} Artificially increase the degree of error locator polynomial for degrees lower than a threshold~\cite{strenzke2008side}.
 
 \item \textbf{To avoid FA:} Check decryption map output weight~\cite{kreuzer2020fault}, re-encrypt, and compare~\cite{kreuzer2020fault} (no countermeasures in place for the attacks mentioned in~\cite{cayrel2020message}).
 
 \item \textbf{To counteract SPA:} Eliminate power trace patterns, branch statements, and data dependencies, maintain consistent power consumption and execution time~\cite{petrvalsky2015countermeasure} (no countermeasures in place for the attacks mentioned in~\cite{guo2022key}).
 
 \item \textbf{To prevent APA:} Employ parallelization, shuffle~\cite{jedlicka2022secure,chen2015horizontal}, or mask the cryptosystem by adding Goppa codewords to the ciphertext during permutation~\cite{jedlicka2022secure,petrvalsky2016differential}.
 
 \item\textbf{For avoiding EM}, no countermeasures in place for the attacks mentioned in~\cite{lahr2020side}.
 
 \item \textbf{To block TMP:} Eliminate the memory access dependency on the content of the lookup-table and achieve constant runtime~\cite{strenzke2008side}.
 
 \item\textbf{For avoiding CB}, no countermeasures in place for the attacks mentioned in~\cite{polanco2019cold}.

\vspace{0.2cm}\end{myBullets}
\end{minipage}} &\med & \med&\med\\ \cline{2-17} 
& \multicolumn{1}{l|}{} & & \multicolumn{1}{l|}{BIKE} & & \multicolumn{1}{l|}{} & \multicolumn{1}{l|}{\doublecheck} & \multicolumn{1}{l|}{\doublecheck} & \multicolumn{1}{l|}{} & \multicolumn{1}{l|}{} & \multicolumn{1}{l|}{} & \multicolumn{1}{l|}{} & &{\begin{minipage}{\linewidth}\begin{myBullets}
\vspace{0.2cm}
 \item \textbf{To prevent TA:} Increase the initial byte generation, eliminate additional random data generation calls, and implement constant-time random number generation techniques~\cite{guo2022don}.

 \item \textbf{To defend against FA:} Employ countermeasures such as default failing, assembly-level instruction duplication, and introducing random delays in the system~\cite{xagawa2021fault} (no countermeasures in place for the attacks mentioned in~\cite{cayrel2020message}).\vspace{0.2cm}\end{myBullets}
\end{minipage}}& \med & \med&\med\\ \cline{2-17}
& \multicolumn{1}{l|}{} & & \multicolumn{1}{l|}{HQC} & & \multicolumn{1}{l|}{} & \multicolumn{1}{l|}{\semidoublecheck} & \multicolumn{1}{l|}{\semidoublecheck} & \multicolumn{1}{l|}{\checkmark} & \multicolumn{1}{l|}{} & \multicolumn{1}{l|}{\doublecheck} & \multicolumn{1}{l|}{} & & {\begin{minipage}{\linewidth}\begin{myBullets}
\vspace{0.2cm}
 \item \textbf{To avoid TA:} Increase the initial byte generation, eliminate additional random data generation calls, and implement constant-time random number generation techniques~\cite{guo2022don}. Also, construct a constant-time decoding algorithm~\cite{wafo2020practicable} (no countermeasures are in place for the attacks mentioned in~\cite{guo2020key}).
 \item \textbf{To defend against FA:} Employ countermeasures such as default failing, assembly-level instruction duplication, and introducing random delays in the system~\cite{xagawa2021fault} (no countermeasures in place for the attacks mentioned in~\cite{cayrel2020message}). 
 \item\textbf{For avoiding TMP}, there are no countermeasures in place for the attacks mentioned in~\cite{schamberger2020power}.
 \item \textbf{To evade EM:} Implement masking using linear secret sharing~\cite{goy2022new}.
\vspace{0.2cm}\end{myBullets}
\end{minipage}}&\high & \med& \high\\ \hline
\multirow{1}{*}{Hash-based} & \multicolumn{1}{l|}{} & & \multicolumn{1}{l|}{} & SPHINCS+ & \multicolumn{1}{l|}{} & \multicolumn{1}{l|}{} & \multicolumn{1}{l|}{\doublecheck} & \multicolumn{1}{l|}{} & \multicolumn{1}{l|}{\doublecheck} & \multicolumn{1}{l|}{} & \multicolumn{1}{l|}{} & & {\begin{minipage}{\linewidth}
\begin{myBullets}
\vspace{0.2cm}
 \item \textbf{To defend against FA:}   (a) Redundant signature computation~\cite{castelnovi2018grafting}, (b) derive few-time signature indices from public values~\cite{castelnovi2018grafting}, (c) establish inter-layer links in the hyper-tree for fault detection~\cite{castelnovi2018grafting}, (d) use sub-tree recomputation with node swaps and enhanced hash function for fault detection~\cite{genet2018practical}, (e) generate and store one-time signatures for efficient reuse~\cite{genet2018practical}, (f) verify vulnerable instructions by recomputing on diverse hardware modules~\cite{genet2018practical}.

 \item \textbf{To prevent APA:} Hide the order of the Mix procedures~\cite{kannwischer2018differential}.\vspace{0.2cm}\end{myBullets}
\end{minipage}}&\med & \med&\med\\ \hline
Isogeny-based & \multicolumn{1}{l|}{} & & \multicolumn{1}{l|}{SIKE} & & \multicolumn{1}{l|}{\checkmark} & \multicolumn{1}{l|}{} & \multicolumn{1}{l|}{{\doublecheck}} & \multicolumn{1}{l|}{} & \multicolumn{1}{l|}{\doublecheck} & \multicolumn{1}{l|}{\doublecheck} & \multicolumn{1}{l|}{} & \multicolumn{1}{l|}{\checkmark} & {\begin{minipage}{\linewidth}\begin{myBullets}
\vspace{0.2cm}
\item\textbf{For avoiding CA}, no countermeasures in place for the attacks mentioned in~\cite{castryck2022efficient}.
\item \textbf{To evade FA:} Employ countermeasures such as default failing, assembly-level instruction duplication, and introducing random delays for variable protection, as well as pushing curves through isogenies and enhancing the probability of recovering correct elliptic curve coefficients during key generation while verifying the implementation afterward~\cite{xagawa2021fault,tasso2021resistance}.
\item \textbf{To counteract APA:} Use of a supersingular curve $E_A$ that generates points of order $3^{e_3}$ in $E_A[3^{e_3}]$~\cite{de2022sike}.
\item \textbf{To prevent EM:} Use of a supersingular curve $E_A$ that generates points of order $3^{e_3}$ in $E_A[3^{e_3}]$~\cite{de2022sike}.
\item\textbf{For avoiding CB}, no countermeasures in place for the attacks mentioned in~\cite{villanueva2020cold}.
\vspace{0.2cm}\end{myBullets}
\end{minipage}}&\high & \med& \high \\ \hline
\end{tabular}}\vspace{3pt}
\footnotesize{$^*$ Attacks and Countermeasures for NIST $4^{th}$ Round PQC Candidates Categorized by Cryptanalysis Attacks (CA), Timing Attacks (TA), Fault Attacks (FA), Simple Power Analysis (SPA), Advanced (correlation/differential) Power Analysis (APA), Electromagnetic Attacks (EM), Template Attacks (TMP) and Cold-Boot Attacks (CB). (\checkmark: Attacks are feasible with no countermeasure in place, \doublecheck: Attacks can be mitigated or prevented by effective  countermeasures, \semidoublecheck: Attacks are viable, but countermeasures are only partially available.)}
\label{tab:side}
\end{table*}
\vspace{-0.2cm}

As discussed earlier, the advent of quantum computing (QC) is poised to revolutionize the landscape of cryptographic algorithm attacks. While the impact of QC on symmetric cryptographic algorithms remains relatively contained, achieved through the adoption of longer keys or extended hash function outputs, public key cryptographic algorithms face a serious threat. This necessitates the replacement of current public key cryptographic algorithms and standards.\\

\begin{enumerate}[wide, font=\itshape, labelwidth=!, labelindent=0pt, label*=\textit{B}.\arabic*.]
\item \textit{Transition to Quantum-Safe Cryptographic Algorithms:}
To safeguard against the emerging QC threat to widely-used public key cryptographic algorithms and transition to a quantum-safe cryptographic environment, the adoption of quantum-safe cryptographic algorithms is imperative. The National Institute of Standards and Technology (NIST) has launched an initiative to standardize quantum-safe cryptographic algorithms, recognizing the vulnerabilities that QC poses to existing cryptographic methods. This initiative encompasses a competition aimed at identifying post-quantum cryptographic algorithms. Post-quantum cryptography encompasses cryptographic algorithms designed to secure Key Exchange (KEM) and Encryption (ENC) and signature (SIG) algorithms against QC-induced threats.

Several categories of post-quantum cryptographic algorithms have emerged, including lattice-based~\cite{bos2018crystals, ducas2018crystals, fouque2018falcon}, code-based~\cite{bernstein2017classic, melchor2018hamming, aragon2017bike}, hash-based~\cite{bernstein2019sphincs+},  and isogeny-based~\cite{jao2011towards} cryptographic algorithms. NIST, cognizant of the QC threat, has taken proactive steps by soliciting post-quantum public-key exchange and digital signature algorithms. In 2022, NIST approved quantum-safe (post-quantum) cryptographic candidates, both for KEM/ENC and Signature, in its fourth round~\cite{NIST_2022July}. These candidates, listed in Table~\ref{tab:side}, should be adopted to ensure quantum-safe cryptography.\\

\item \textit{Challenges Beyond Quantum-Resistant Algorithms:}  
NIST's Post-Quantum Cryptography (NIST PQC) competition endeavors to establish new cryptographic standards that can withstand QC attacks. However, it is essential to recognize that even post-quantum secure cryptographic algorithms may still be susceptible to other types of attacks, such as side-channel and cryptanalysis attacks.

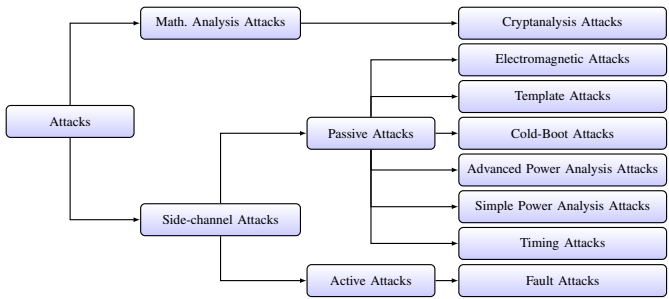
\begin{figure}[!htbp]
\centering
\large
\resizebox{\linewidth}{!}{
\begin{tikzpicture}[line width=0.035cm]
\node (d0) [draw,top color=white, bottom color=blue!20, rounded corners,minimum height=1cm,minimum width=4cm] at (0,0) {Attacks};
\node (d1) [draw,top color=white, bottom color=blue!20, rounded corners,minimum height=1cm,minimum width=5cm, above right of=d0,xshift=4cm,yshift=2.37cm] {Math. Analysis Attacks};

\node (d2) [draw,top color=white, bottom color=blue!20, rounded corners,minimum height=1cm,minimum width=5cm, below right of=d0,,xshift=4cm,yshift=-2.37cm] {Side-channel Attacks};

\node (d21) [draw,top color=white, bottom color=blue!20, rounded corners,minimum height=1cm,minimum width=4cm,below right of=d2,xshift=4cm,yshift=-1.2cm] {Active Attacks};
\node (d211) [draw,top color=white, bottom color=blue!20, rounded corners,minimum height=1cm,minimum width=6.5cm, right of=d21,xshift=5cm] {Fault Attacks};

\node (d22) [draw,top color=white, bottom color=blue!20, rounded corners,minimum height=1cm,minimum width=4cm, above right of=d2,
 ,xshift=4cm,yshift=2cm] {Passive Attacks};
 \node (d223) [draw,top color=white, bottom color=blue!20, rounded corners,minimum height=1cm,minimum width=6.5cm, right of=d22,xshift=5cm,,yshift=2.3cm] {Electromagnetic Attacks};
 \node (d222) [draw,top color=white, bottom color=blue!20, rounded corners,minimum height=1cm,minimum width=6.5cm, right of=d22,xshift=5cm,,yshift=1.15cm] {Template Attacks};
 \node (d221) [draw,top color=white, bottom color=blue!20, rounded corners,minimum height=1cm,minimum width=6.5cm, right of=d22,xshift=5cm,,yshift=0cm] {Cold-Boot Attacks};
\node (d224) [draw,top color=white, bottom color=blue!20, rounded corners,minimum height=1cm,minimum width=6.5cm, right of=d22,xshift=5cm,,yshift=-1.15cm] { Advanced Power Analysis Attacks};
\node (d225) [draw,top color=white, bottom color=blue!20, rounded corners,minimum height=1cm,minimum width=6.5cm, right of=d22,xshift=5cm,,yshift=-2.3cm] {Simple Power Analysis Attacks};
\node (d226) [draw,top color=white, bottom color=blue!20, rounded corners,minimum height=1cm,minimum width=6.5cm, right of=d22,xshift=5cm,,yshift=-3.45cm] {Timing Attacks};
 \node (d11) [draw,top color=white, bottom color=blue!20, rounded corners,minimum height=1cm,minimum width=6.5cm,xshift=9.7cm, right of=d1] {Cryptanalysis Attacks};

 \draw[-latex] (d0) |- (d2);
 \draw[-latex] (d0) |- (d1);
 \draw[-latex] (d2) |- (d21);
 \draw[-latex] (d2) |- (d22);
 \draw[-latex] (d1) -- (d11);
 \draw[-latex] (d21) -- (d211);
 \draw[-latex] (d22) -- (d221); 
 \draw[-latex] (d22) |- (d222);
 \draw[-latex] (d22) |- (d223);
 \draw[-latex] (d22) |- (d224);
 \draw[-latex] (d22) |- (d225); 
 \draw[-latex] (d22) |- (d226); 
\end{tikzpicture}}

\caption{Taxonomy of Attacks for Quantum-Safe Cryptographic Candidates for Standardization (i.e., NIST $4^{th}$ Round PQC Candidates)}
\label{fig:Side}
\end{figure}

A side-channel attack exploits information leakage during the execution of a cryptographic algorithm, including power consumption, electromagnetic radiation, or timing information, as shown in Figure~\ref{fig:Side}. By analyzing this leaked information, an attacker may extract sensitive data, such as a private key. Cryptanalysis attacks, on the other hand, aim to break the encryption or signature schemes by identifying structural weaknesses in the algorithm. 
Significant instances of side-channel and cryptanalysis attacks on NIST's fourth-round candidates have been reported. Notably, the evaluation process is ongoing, and more attacks may emerge in the future. This section examines these attacks, potential countermeasures, and the associated threats for post-quantum cryptographic algorithms considered by NIST as quantum-safe cryptographic candidates.\\

\item \textit{Quantum Attack Vectors for Quantum-Safe Cryptography:} 
A quantum attacker may persistently attempt to compromise post-quantum cryptography (PQC) by exploiting various vulnerabilities, including side-channels or mathematical analysis. Figure~\ref{fig:Side} provides a taxonomy of these vulnerabilities that a quantum attacker could exploit to breach the security of post-quantum cryptographic algorithms considered for NIST standardization in the fourth round.

In light of this, an attacker can execute side-channel or mathematical analysis-type attacks. Consequently, we assess the risks associated with each attack on the candidates listed in Table~\ref{tab:side}. Our evaluation hinges on a qualitative risk assessment, which factors in both likelihood and impact, as visualized in Figure~\ref{fig:risk-matrix}.\\

\item \textit{Likelihood and Impact Evaluation:} 
In the likelihood assessment, we examine various factors such as exploitability (via physical access, network, or the internet), available countermeasures (detailed in Table~\ref{tab:side}), and the criteria provided in Appendix G of NIST-SP 800-30~\cite{NIST-SP800-30}. We categorize likelihood levels as follows:

\begin{enumerate}
\item If a known exploit is available and can be launched from the Internet or a network, the likelihood is considered high when no countermeasures are in place; otherwise, it is assessed as medium.
\item When a known exploit exists, there are no countermeasures, and physical access to the target system is required, the likelihood is rated as medium.
\item In cases where no known exploit exists (or exists along with an effective countermeasure), and a malicious user necessitates administrative or elevated privileges to execute an attack on the target system, the likelihood is determined to be low.
\end{enumerate}

For impact assessment, we refer to the criteria outlined in Appendix H of NIST-SP 800-30~\cite{NIST-SP800-30}. Threats stemming from quantum attackers have the potential to significantly impact user satisfaction, expose personal data and sensitive information, or damage an organization's reputation. Therefore, the impact level is considered medium.

In our research, we provide a comprehensive analysis of the probability and impact levels, along with the associated risks for various post-migration algorithms. We summarize our findings in Table~\ref{tab:side}, which includes details on vulnerabilities, potential attacks, countermeasures, and post-quantum threats associated with different post-migration algorithms.
\end{enumerate}
\section{Cybersecurity Implications of Quantum Computing on Digital infrastructure in Pre-Migration Era}\label{sec:pre}

Digital infrastructure, crucial to contemporary society, is structured across nine vital service layers: applications, data, runtime, middleware, operating systems, virtualization, hardware, storage, and networks. These layers work in unison to support global Internet-based services. Our research examines the impact of classical cryptography on these layers, identifying and addressing the threats posed by the advent of quantum computing to both existing and upcoming digital ecosystems, including cloud platforms.

Focusing on what we term the Pre-Migration Phase, it is essential to analyze the current landscape. This phase zeroes in on the Cyber Impact of Quantum Computing on infrastructures that currently depend on classical cryptography. With the progression of quantum computing technologies, existing cryptographic standards, which are pivotal for the protection of infrastructure and confidential data, are increasingly at risk of being breached. The delay in adopting quantum-resistant algorithms may expose organizations to a spectrum of quantum-enabled cyberattacks. We explore how advances in quantum computing, including developments in quantum engineering and cryptography, jeopardize classical cryptographic systems, underlining the profound risks these advancements pose on conventional encryption techniques. 
\subsection{Cyber Impact of Quantum Computing in the Pre-Migration Phase}

The advent of quantum computing presents significant security challenges to traditional cryptographic systems. Many current cryptographic algorithms, crucial for safeguarding infrastructure and confidential data, are at risk of being rendered ineffective against quantum computer attacks. Organizations lagging in the adoption of quantum-resistant cryptographic methods are exposed to several potential quantum-enabled threats, including (a) cryptographic breaches, (b) identity theft, (c) financial fraud, and (d) data tampering.\\

\begin{enumerate}[wide, font=\itshape, labelwidth=!, labelindent=0pt, label*=\textit{A}.\arabic*.]
 \item \textit{Cryptographic Breaches:}  Quantum computers have the capability to decrypt standard encryption methodologies such as RSA, Diffie-Hellman, and elliptic curve cryptography. This breach could lead to unauthorized access to critical information, encompassing private communications, passwords, and financial transactions.\\

 \item \textit{Identity Theft:} The exploitation of digital signatures by quantum computers can facilitate identity theft. Impersonation of legitimate users or entities could result in unauthorized access to high-security systems or sensitive data, impacting governmental or military operations.\\

 \item \textit{Financial Fraud:} The ability of quantum computers to undermine cryptographic security in financial transactions could lead to the misappropriation of funds, unauthorized transfers, or manipulation of financial records.\\

 \item \textit{Data Tampering:} Quantum computing might enable alterations in digital data storage, leading to potential manipulations in medical records, financial statements, or electoral databases.\\

 \item \textit{Cyber Espionage:}  Nation-states or powerful entities harnessing quantum computing capabilities could engage in advanced espionage activities, targeting confidential and strategic data such as industrial trade secrets or national security information.
\end{enumerate}

To mitigate these vulnerabilities, organizations are advised to proactively transition to quantum-resistant cryptographic technologies. In addition to this cryptographic upgrade, implementing robust access controls and advanced threat detection and response mechanisms are essential strategies for reinforcing cybersecurity in the quantum era.

\subsection{Quantum Computing Attack Vectors in the Pre-Migration Phase}

Classical systems are affected by quantum computers because it has many potential uses, such as quantum engineering, cryptography, machine learning, and artificial intelligence. As a result, a quantum attacker has the ability to break some of the most prominent encryption algorithms in use today. Therefore, this section surveyed existing security challenges present in the infrastructure before migration to the post-quantum cryptography standard and presented the threats, vulnerabilities, attack vectors, and cryptography types of existing classical systems for evaluating the risk that quantum attackers can cause. This classification can assist security analysts in discovering mitigation measures and identifying security issues more quickly. The attack vectors that can be caused by quantum machines are discussed below.\\

\begin{enumerate}[wide, font=\itshape, labelwidth=!, labelindent=0pt, label*=\textit{B}.\arabic*.]
\item \textit{Code Injection:} In the context of quantum computing, attackers might leverage the advanced capabilities of quantum algorithms, such as Shor's algorithm for rapid factorization and Grover's algorithm for efficient database search, to undermine cryptographic security. These quantum algorithms could potentially break current encryption methods, enabling attackers to forge digital signatures. This vulnerability might allow the insertion of malicious code into software updates or trusted communications.
Potential threats include Application Software Exploitation~\cite{compastie2020virtualization, hoffman2020web}, Insecure Code~\cite{owasp_code_injection, hoffman2020web}, Keyloggers~\cite{commomtypmalware}, Viruses~\cite{ThreatsProtection}, Trojan Horses~\cite{khan2017stride}, Worms~\cite{rawal2023malware}, Rootkits~\cite{rawal2023malware}, Trap Doors~\cite{TrapdoorVIRUS}, Crimeware~\cite{khan2017stride}, Fileless Malware~\cite{commomtypmalware}, Data Theft~\cite{Threatinsiderdef}, and System Sabotage~\cite{Threatinsiderdef}. Such attacks could compromise the integrity, confidentiality, and availability of systems and data.\label{code-injection}\\

\item \textit{Hypervisor Exploitation:} In virtualized environments, a significant threat emerges from quantum attackers targeting the hypervisor. Leveraging the advanced capabilities of quantum algorithms, these attackers might successfully bypass established cryptographic defenses to inject malicious code. This intrusion can result in elevated privileges, granting the attackers overarching control over the virtualized framework~\cite{compastie2020virtualization}. The consequences are severe, including data interception or manipulation and the unauthorized exfiltration of sensitive information. Furthermore, such breaches can disrupt the functionality of all dependent virtual machines, increasing the risk of a full-scale network breakdown. These vulnerabilities might also be exploited to orchestrate extensive denial-of-service (DoS) attacks, further amplifying their destructive impact.\\

\item \textit{VM Migration and Quantum Risks:} In the process of Virtual Machine (VM) migration, sensitive data is generally encrypted to ensure security while in transit~\cite{oberheide2008empirical}. Nevertheless, the advent of quantum computing presents a significant threat, as it equips attackers with the ability to rapidly decrypt this information. This vulnerability could lead to unauthorized access to confidential data and potentially compromise critical services, including cloud-based operations and data center functionalities.\label{VM-Migration}\\

 \item \textit{OS Kernel Exploitation:} Quantum attackers could target the core of a computer's operating system - the kernel. By leveraging algorithms like Shor's and Grover's, they can undermine the cryptographic safeguards protecting kernel integrity and authentication. A successful attack here could lead to a complete system takeover, data leaks, and the introduction of deep system vulnerabilities~\cite{compastie2020virtualization}.\\

 \item \textit{Firmware Exploitation:} Firmware, frequently underestimated in cybersecurity strategies, emerges as a pivotal point of vulnerability against quantum-enabled adversaries. Such attackers could leverage advanced quantum algorithms, notably Shor's and Grover's, to undermine cryptographic safeguards of firmware updates. This vulnerability permits the unauthorized insertion of malicious firmware across a spectrum of devices, extending from IoT appliances to vital infrastructure systems. The resultant risks span espionage, strategic sabotage, and potentially widespread operational disruptions~\cite{cuozzo2016critical}.\\

 \item \textit{Symmetric Cryptographic Key Recovery:} Leveraging quantum algorithms like Grover's, quantum attackers can significantly reduce the time required to brute-force symmetric cryptographic keys. This poses a grave threat to data confidentiality and can lead to unauthorized access and data breaches~\cite{brassard1997quantum,grover1996fast,grover2020efficient}.\label{grover-item}\\

 \item \textit{Unauthorized Data Exfiltration:} The advent of quantum computing introduces significant security challenges, especially concerning the safety of encrypted data in virtual machines (VMs) and during data transmission. Quantum algorithms like Shor's and Grover's threaten to undermine traditional cryptographic defenses. These quantum algorithms enable attackers to potentially decrypt and access confidential data within VMs and encrypted communications, bypassing existing encryption techniques. Such breaches have profound implications for individual privacy, corporate confidentiality, and national security. The main tactics for this type of data exfiltration encompass Inter-VM Communication Introspection~\cite{compastie2020virtualization}, Decrypting Encrypted Network Traffic~\cite{brassard1997quantum, grover1996fast, shor1994algorithms}, Man-In-The-Middle (MITM) Attacks~\cite{chakraborty2017minimum}, Enhanced Cryptanalysis~\cite{brassard1997quantum, grover1996fast, shor1994algorithms}, Cross-VM Side-channel Cache Attacks utilizing Cryptanalysis~\cite{gruss2019page, esfahani2021enhanced}, and Software-Induced Fault Attacks~\cite{krautter2018fpgahammer}, among others.\label{Exfiltration}\\

 \item \textit{Storage Server/Disk Encryption Breach:} Quantum attackers are capable of breaching storage encryption by illicitly decrypting keys using sophisticated algorithms, such as Shor's and Grover's. This vulnerability endangers not just the confidentiality but also the integrity of stored data. Such intrusions could lead to unauthorized access, alteration, or corruption of vital information, thereby posing a serious threat to the privacy and reliability of the data~\cite{gort2022relational,jangjou2022comprehensive}.\label{storage}\\

\item \textit{POST Request Collision:} Quantum-enhanced collision attacks can be executed by overwhelming servers with POST requests that are particularly resource-intensive to process. These attacks leverage the quantum computational advantage to exacerbate the challenges in handling POST requests, typically causing server overloads, crashes, or significant performance degradation, thus undermining service availability. In such scenarios, a quantum attacker can dispatch multiple voluminous POST messages with dynamically generated, hash-intensive payloads. This approach significantly increases the demand for hash computation, consequently heightening the risk and impact of a collision attack~\cite{smith2019postcollision}.\label{post}\\

\item \textit{Quantum-Enhanced Encryption Exploitation:} 
Quantum attackers, wielding significantly more computational power, can encrypt data using complex algorithms far beyond the capabilities of classical computing. This advanced encryption can be maliciously applied in forms such as quantum-enhanced ransomware, making the encrypted content incredibly difficult, if not impossible, to decrypt with conventional methods~\cite{craciun2019trends,kilber2021cybersecurity}.\label{ransomeware} \\

\item \textit{Access Points (AP) Breaches:} Wi-Fi networks and other network access points secured with legacy protocols like WPA or WEP are at risk. These protocols depend on cryptographic algorithms vulnerable to quantum computing attacks. Specifically, Shor’s algorithm can break cryptosystems based on public-key cryptography, which is a foundation of many current security protocols, while Grover's algorithm poses a threat to systems using symmetric key cryptography. Brassard et al.’s contributions in quantum computing further indicate vulnerabilities in these older security protocols~\cite{brassard1997quantum}. Such quantum computing advancements could enable attackers to gain unauthorized access by decrypting encrypted communications, leading to interception or network-based attacks~\cite{lounis2020attacks}.\label{access}\\

\end{enumerate}

\subsection{Security Challenges in each Infrastructure Layer}

\begin{table*}[!htbp]
\caption{Pre-Migration Security Challenges for Each Infrastructure Layer}
\large
\renewcommand{\arraystretch}{1.2}
\resizebox{\linewidth}{!}{%
\begin{tabular}{|p{0.15\linewidth}|p{0.2\linewidth}|p{0.25\linewidth}|p{0.3\linewidth}|p{0.48\linewidth}|p{0.67\linewidth}|p{0.35\linewidth}|}
\hline 
\textbf{Layer} & \textbf{STRIDE Threat} & \textbf{Vulnerabilities} & \textbf{Exploited by} & \textbf{Attack Vector} & \textbf{Countermeasures} & \textbf{Responsibility} \\ \hline
\multirow{17}{2cm}{Application}
& Spoofing 
& Weak Encryption Algorithms 
& VM Mobility~\cite{compastie2020virtualization,huang2012security}, Brute Force Attack~\cite{merhav2019universal} 
& Quantum algorithms like Shor's and Grover's could be used to decrypt VMs during migration or brute force cryptographic keys 
& Transition to lattice-based or other NIST-approved post-quantum cryptographic standards, strengthen key management practices 
& Software vendors, Security architects \\
\cline{2-7}
& Tampering 
& Insecure Code 
& Application Software Exploitation~\cite{compastie2020virtualization,hoffman2020web} 
& Code injection with forged signature during software update, exploiting vulnerabilities in the update process 
& Adoption of secure coding standards, Code audits, Implement post-quantum secure signatures 
& Application developers, Security teams \\
\cline{2-7}
& Repudiation 
& Weak Authentication Mechanisms 
& Assertion Repudiation including Hash Attack, Data Driven Attacks and Impersonation/Masquerade Attack~\cite{grassi2020digital,aldeco2012secure} 
& Quantum algorithms can forge signatures or hashes used in authentication 
& Deployment of quantum-resistant authentication mechanisms, Use of hardware security modules for key storage 
& Service providers, End-users \\

\cline{2-7}
& Info. Disclosure 
& Insufficient Data Protection 
& Application Software Exploitation~\cite{compastie2020virtualization,hoffman2020web}, Inter-VM Communication Introspection~\cite{compastie2020virtualization}, Cryptanalysis Attacks~\cite{brassard1997quantum,grover1996fast,shor1994algorithms} 
& Code injection during updates, MITM attacks on VMs, cryptographic breaches 
& Implement quantum-safe encryption for data-at-rest and in-transit, utilize secure key exchange protocols 
& Service providers, Network administrators \\
\cline{2-7}
& Denial of Service 
& Vulnerable Network Interfaces 
& POST Request Collision~\cite{smith2019postcollision} & Quantum-enhanced collision attacks causing server overloads 
& Load balancing, Deployment of quantum-resistant hash functions, real-time traffic monitoring 
& Infrastructure operators, Network engineers \\
\cline{2-7}

& Elevation of Privilege 
& Flawed Permission Models 
& Application Software Exploitation~\cite{compastie2020virtualization,hoffman2020web} 
& Code injection leading to unauthorized privilege escalation 
& Enforce robust access control policies, Regular privilege audits, Zero trust architecture implementation 
& System administrators, IT departments \\
\hline
\multirow{12}{*}{Data} 
& Spoofing 
& Susceptible encryption under quantum attacks 
& Brute Force Attack~\cite{gort2022relational} 
& Quantum algorithms reducing symmetric key security 
& Transition to quantum-resistant encryption standards; promote best practices in cryptographic best practices.
& Industry Standards Bodies, Security Architects. \\ \cline{2-7} 

& Tampering 
& Modification of encrypted data storage 
& Data Modification Attack~\cite{jangjou2022comprehensive}
& Storage servers encryption breach by quantum attacker
& Regular integrity checks; integrate quantum-resistant cryptographic hashes and encryption.
& Data Storage Providers, Security Engineers.\\ \cline{2-7} 

& Info. Disclosure 
& Insecure software update mechanisms 
& Injecting Keylogger~\cite{commomtypmalware} 
& Code injection via quantum-forged signatures 
& Enhance update mechanisms with quantum-resistant signatures; secure distribution networks with robust authentication protocols.
& Software Vendors, System Administrators. \\ \cline{3-7} 

& 
& Classical cryptosystems breakable by quantum algorithms 
& Cryptanalysis Attacks~\cite{brassard1997quantum,grover1996fast,shor1994algorithms} 
& Quantum decryption via Shor's, Grover's and Brassard et al.'s algorithms 
& Early adoption of NIST-approved post-quantum algorithms; update encryption protocols.
& Cryptographers, Software Developers \\ \cline{2-7} 

& Denial of Service 
& Quantum-enhanced collision attacks on hash functions 
& Hash DoS~\cite{masdari2016survey}
& POST request collision using quantum computing 
& Implement rate limiting; adopt post-quantum hash functions; use advanced intrusion detection systems.
& Network Administrators, Cybersecurity Teams \\ \cline{2-7} 

& Elevation of Privilege 
& Weak cryptographic access controls 
& Data Theft~\cite{Threatinsiderdef} 
& Quantum algorithms enabling unauthorized access 
& Employ multi-factor authentication; establish strict access control policies using post-quantum secure methods. 
& IT Security Policy Makers, End-Users \\ \cline{3-7} 
& 
& Compromised cryptographic mechanisms 
& Unauthorized Data Access~\cite{jangjou2022comprehensive} 
& Quantum cryptanalysis for access control breach 
& Upgrade to quantum-resistant access control systems; continuous security training for users.
& Access Control Vendors, Corporate Training Departments \\ 
\hline
\multirow{12}{*}{Runtime} &
\multirow{3}{*}{Spoofing} & Susceptibility to quantum decryption & {VM Mobility}~\cite{compastie2020virtualization,huang2012security} &
Decryption of an encrypted VM while moving it from one physical host to another & Implement quantum-resistant encryption standards; Employ hybrid encryption models & Vendors, Security Architects \\ \cline{3-7} 
& 
& Weak encryption methods 
& Brute Force Attack~\cite{tirado2018new} 
& Via Grover's algorithm to brute-force symmetric cryptographic keys 
& Early adoption of post-quantum cryptographic algorithms; Regular security updates 
& Implementers, IT Security Teams \\ \cline{2-7}

& \multirow{1}{*}{Tampering} 
& Insecure update mechanisms 
& Application Software Exploitation~\cite{compastie2020virtualization,hoffman2020web} 
& Code injection with forged signature during software update 
& Deployment of quantum-safe digital signature schemes; Use of secure and authenticated update channels 
& Vendors, Developers\\ \cline{2-7}

& Info. Disclosure 
& Inadequate encryption
& Application Software Exploitation~\cite{compastie2020virtualization,hoffman2020web}
& Code injection leading to data leaks
& Transition to quantum-resistant encryption protocols; Use of secure communication channels 
& Vendors, Network Administrators \\ 
\cline{3-7} 
& 
& Vulnerable VM communication 
& Inter-VM Communication Introspection~\cite{compastie2020virtualization}
& MITM attacks exploiting quantum decryption capabilities 
& Establishment of quantum-secure channels; Regular penetration testing 
& Implementers, Security Teams \\ \cline{3-7}
& 
& Weak cryptosystems 
& Cryptanalysis Attacks~\cite{brassard1997quantum,grover1996fast,shor1994algorithms} 
& Data exfiltration using quantum algorithms to break encryption
& Upgrading to quantum-resistant algorithms; Conducting regular security audits 
& Vendors, CISOs \\ \cline{2-7} 
& {Elevation of Privilege} 
& Compromised update integrity
& Application Software Exploitation~\cite{compastie2020virtualization,hoffman2020web} 
& Forged updates granting elevated access.
& Secure bootstrapping and update processes; Multi-factor authentication systems 
& Implementers, Users, IT Administrators \\ 
\hline 
\multirow{14}{*}{Middleware} 
&Spoofing 
& VM Identity Forgery 
& VM Mobility~\cite{compastie2020virtualization,huang2012security} 
& Decryption of an encrypted VM during migration 
& Implementation of quantum-resistant cryptographic protocols for secure VM migration
& Vendors to provide tools; Implementers to deploy\\ \cline{3-7} 
 &
 &Cryptographic Key Cracking 
 & Brute Force Attack~\cite{bates2012detecting}
 & Quantum algorithms like Grover's 
 & Adoption of quantum-resistant public key infrastructures (PKIs) 
 & Vendors to update products; Implementers to integrate \\ \cline{2-7}
 & Tampering 
 & Code 
 & C\&C Channel Exploitation~\cite{compastie2020virtualization} 
 & MITM attacks with forged signatures 
 & Usage of stateful hash-based signatures for code and communications 
 & Vendors to develop; Implementers to apply \\ \cline{3-7}
 &  
 & Hypervisor Integrity 
 & Hypervisor Exploitation~\cite{compastie2020virtualization} 
 & Direct attacks on the hypervisor 
 & Deployment of verified boot processes and attestation mechanisms 
 & Vendors to innovate; Implementers to maintain\\ \cline{2-7} 
 & Repudiation 
 & Non-repudiation Failure 
 & C\&C Channel Exploitation~\cite{compastie2020virtualization} 
 & Forging signatures on C\&C channels 
 & Immutable logging mechanisms with quantum-resistant signatures
 & Implementers to enforce; Users to monitor\\
 \cline{2-7} 
 & Info. Disclosure 
 & Inter-VM Communication Eavesdropping 
 & Cross-VM Side-channel Cache Attacks~\cite{shahzad2015virtualization,anwar2017cross,saeed2020cross} 
 & Cryptanalysis with quantum algorithms 
 & Isolation enhancements with quantum-safe encrypted channels 
 & Vendors to enhance; Implementers to configure\\ \cline{2-7} 
 & Denial of Service 
 & Hypervisor Overload 
 & Hypervisor Exploitation~\cite{compastie2020virtualization} 
 & Quantum DoS attacks via hypervisor exploitation 
 & Redundant, distributed architectures and quantum-aware rate limiting 
 & Vendors to design; Implementers to orchestrate\\ \cline{2-7} 
 & Elevation of Privilege
 & Unauthorized Access 
 & Hypervisor Exploitation~\cite{compastie2020virtualization} 
 & Gaining higher privileges via hypervisor exploitation 
 & Multifactor authentication systems with quantum-resistant protocols 
 & Implementers to enforce; Users to adopt 
 \\ \hline
\multirow{40}{*}{OS} 
&{Spoofing} 
& Quantum Algorithm Exploits 
& Injecting Virus~\cite{ThreatsProtection} 
& Code injection with forged signature during software update 
& Implementation of Quantum-Secure Signature Schemes, Transition to Quantum-Resistant Certificates 
& OS Developers, System Administrators \\ \cline{3-7} 
& 
& Cryptographic Breakdown 
& Injecting Insecure Code~\cite{owasp_code_injection,hoffman2020web} 
& Code injection with forged signature during software update 
& Adoption of Post-Quantum Cryptography Libraries, Secure Software Development Lifecycle (SDLC) Practices 
& Software Vendors, Security Teams \\ \cline{4-7}
& 
& 
& Injecting Trap Door~\cite{TrapdoorVIRUS}
 & Malicious code, containing a trap door, which is injected into software with a forged signature during software update. 
 & Enhanced Software Supply Chain Security, Continuous Integrity Monitoring 
 & System Administrators, Security Auditors \\ \cline{3-7} 
 
 & 
 & Symmetric Key Exposure 
 & Brute Force Attack~\cite{tirado2018new} 
 & Exploitation of symmetric key cryptography via Grover's algorithm 
 & Migration to Larger Symmetric Key Sizes, Quantum Key Distribution (QKD) Integration 
 & Cryptographers, Security Architects \\ \cline{2-7} 
 & {Tampering} 
 & Kernel Integrity Compromise & 
 OS Kernel Exploitation~\cite{compastie2020virtualization}
 & Exploitation through breaking hardware security modules (HSM) using quantum algorithms. 
 & Secure Kernel Development Techniques, Quantum-Resistant HSMs 
 & OS Vendors, Hardware Providers \\ \cline{3-7} 
 & 
 & Malware Insertion 
 & Injecting Virus~\cite{ThreatsProtection} 
 & Code injection with forged signature during software update 
 & Advanced Endpoint Detection and Response (EDR) Tools, Regular Security Updates 
 & IT Security Teams, End Users\\ \cline{4-7} 
 & 
 & 
 & Injecting Insecure Code~\cite{owasp_code_injection,hoffman2020web} 
 & Code injection with forged signature during software update. 
 & Mandatory Access Control (MAC), Executable Space Protection 
 & System Administrators, Application Developers \\ \cline{4-7} 
 & 
 & 
 & Injecting Trojan Horse~\cite{khan2017stride}
 & Code injection with forged signature during software update. 
 & Real-Time Threat Intelligence, Heuristic Analysis Techniques & IT Security Analysts, End Users \\ \cline{4-7} 
 & 
 & 
 & Crimeware~\cite{khan2017stride} 
 & Code injection with forged signature during software update. 
 & Comprehensive Cybersecurity Training, Security Operations Center (SOC) 
 & IT Staff, SOC Analysts\\ \cline{2-7} 
 
 &\multirow{2}{*}{Repudiation} 
 & Non-repudiation Failure 
 & Injecting Virus~\cite{ThreatsProtection}, Injecting Insecure Code~\cite{owasp_code_injection,hoffman2020web} 
 & Code injection with forged signature during software update 
 & Implement quantum-resistant cryptographic solutions for signature verification; Enhance audit and logging mechanisms to track software changes and updates; Utilize multi-factor authentication to reinforce non-repudiation.
 & Vendors to provide quantum-safe signature schemes; Implementers to ensure proper configuration and monitoring; Users to adhere to security practices. \\ \cline{2-7}
 
 &{Info. Disclosure} 
 & Quantum Cryptanalytic Attacks 
 & OS Kernel Exploitation~\cite{compastie2020virtualization} 
 & Exploited through breaking HSM via Shor's, Grover's, and Brassard et al.'s algorithms by quantum attackers. 
& Implementing hardware-enforced security measures that remain unaffected by quantum decryption capabilities, such as Trusted Platform Modules (TPMs) for secure boot; Integrating quantum-resistant hardware security features; Regular kernel integrity checks using cryptographic hashes based on post-quantum algorithms.
& OS Vendors, Hardware Manufacturers \\ \cline{3-7} 
 &
 & Weaknesses in Encrypted Data Handling 
 & Inter-VM Communication Introspection~\cite{compastie2020virtualization} 
& Man-in-the-Middle (MITM) and data exfiltration of encrypted content by quantum-capable adversaries.
& Transition to Quantum-Secure Channels, Real-time monitoring and anomaly detection systems.
& Implementers to enforce new protocols; Security teams to monitor and respond to incidents. \\ \cline{3-7} 
 & 
 & Malicious Code Insertion 
 & Injecting Insecure Code~\cite{owasp_code_injection,hoffman2020web}, Keylogger~\cite{commomtypmalware}, Trojan Horse~\cite{commomtypmalware}, Fileless Malware~\cite{commomtypmalware}, Worm~\cite{rawal2023malware} 
 & Code injection with forged signature during software update 
 & Implement quantum-resistant cryptographic algorithms for code signing, advanced threat detection systems with heuristic analysis, secure software update mechanisms, employ SSDLC, and conduct regular developer training on quantum-related threats. & Developers adhere to SSDLC with quantum-resistant measures, vendors provide quantum-resistant code signing tools, and users verify signatures with quantum-resistant verification methods. \\ \cline{3-7} 
 & & Cross-VM Side-channel Attacks 
 & Side-channel Cache Attacks in combination with cryptanalysis~\cite{gruss2019page,esfahani2021enhanced} & Data exfiltration via Shor's, Grover's, and Brassard et al.'s algorithms, leveraging side-channel vulnerabilities.
& Strengthening VM isolation, employing noise and blinding techniques in cryptographic processes, and regular security audits for virtualized environments.
& Virtualization Platform Providers, System Administrators  
\\ \hline

\end{tabular}
}
\end{table*}

\addtocounter{table}{-1}
 
\begin{table*}[!htbp]
\caption{(Cont.) Pre-Migration Security Challenges for Each Infrastructure Layer}
\large
\renewcommand{\arraystretch}{1.2}
\resizebox{\linewidth}{!}{%
\begin{tabular}{|p{0.15\linewidth}|p{0.2\linewidth}|p{0.25\linewidth}|p{0.3\linewidth}|p{0.48\linewidth}|p{0.67\linewidth}|p{0.35\linewidth}|}
\hline 
\textbf{Layer} & \textbf{STRIDE Threat} & \textbf{Vulnerabilities} & \textbf{Exploited by} & \textbf{Attack Vector} & \textbf{Countermeasures} & \textbf{Responsibility} \\ \hline

 \multirow{15}{*}{OS} 
 & {Denial of Service} 
 & Malicious code inserted via signature forgery 
& Injecting Virus~\cite{ThreatsProtection} 
& Code injection with forged signature during software updates 
& Quantum-secure code signing protocols; Advanced heuristic-based antivirus solutions; Secure software update mechanisms 
& Application Developers, End Users \\ \cline{3-7} 

 & & Quantum-Enhanced Encryption & Ransomware~\cite{craciun2019trends,kilber2021cybersecurity} & Triggered when more powerful quantum attackers have more power to encrypt contents in a more complex way. & Backup systems, Anti-ransomware solutions & Users, Implementers \\ \cline{2-7} 
& Elevation of Privilege 
& Kernel integrity compromised by quantum algorithms & OS Kernel Exploitation~\cite{compastie2020virtualization} 
& Bypassing hardware security modules (HSMs) via quantum algorithms (Shor's, Grover's and Brassard et al.'s algorithms) 
& Implementing hardware-enforced security measures that remain unaffected by quantum decryption capabilities, such as Trusted Platform Modules (TPMs) for secure boot; Integrating quantum-resistant hardware security features; Regular kernel integrity checks using cryptographic hashes based on post-quantum algorithms.
& OS Vendors, Hardware Manufacturers\\ \cline{3-7} 
& & Malicious code inserted via signature forgery 
& Injecting Virus~\cite{ThreatsProtection}, Insecure Code~\cite{owasp_code_injection,hoffman2020web}, Trap Door~\cite{TrapdoorVIRUS} 
& Code injection with forged signature during software updates 
& Implement quantum-resistant cryptographic algorithms for code signing, advanced threat detection systems with heuristic analysis, secure software update mechanisms, employ SSDLC and conduct regular developer training on quantum-related threats. & Developers adhere to SSDLC with quantum-resistant measures, vendors provide quantum-resistant code signing tools, and users verify signatures with quantum-resistant verification methods. 
\\ \hline

 \multirow{12}{*}{Virtualization} 
 & \multirow{1}{*}{Spoofing} 
& Cryptographic key vulnerability
& Brute Force Attack~\cite{grover2020efficient}
& Via Grover's algorithm to brute-force a key
& Upgrade to cryptographic algorithms with larger key sizes, and encourage early adoption of post-quantum cryptographic standards.
& Vendors to develop, Implementers to deploy \\ \cline{2-7}
& Info. Disclosure
& Hypervisor vulnerabilities
& Side-channel cache attacks~\cite{irazoqui2015s} and cryptanalysis attack ~\cite{brassard1997quantum,grover1996fast,shor1994algorithms}
& Data exfiltration via quantum algorithms
& Introduce hardware-enforced isolation, leverage intrusion detection systems to monitor anomalous activities, and apply quantum-resistant algorithms for sensitive operations.
& Vendors to innovate, Implementers to install and maintain \\ \cline{3-7}
& & Cross-VM attacks
& Cross-VM Side-channel Cache Attacks~\cite{shahzad2015virtualization,anwar2017cross,saeed2020cross} in combination  with Cryptanalysis~\cite{brassard1997quantum,grover1996fast,shor1994algorithms}
& Data exfiltration via quantum algorithms
& Segregate critical VMs, implement strict access control and continuous monitoring to detect and respond to side-channel attacks.
& Implementers to configure, IT security to oversee\\ \cline{3-7}
& & VM migration risks
& VM Migration~\cite{oberheide2008empirical}
& Data exfiltration during VM migration
& Employ end-to-end encryption for VM migration data, utilize secure migration protocols, and conduct regular security audits on migration processes.
& IT security to audit, Implementers to execute\\ \hline
\multirow{9}{*}{Hardware} &
Spoofing 
& Quantum key-breaking 
& Brute Force Attack~\cite{lu2019attacking} 
& Exploiting quantum algorithms to break cryptographic keys 
& Implement quantum-resistant cryptographic algorithms 
& Vendors \\ \cline{2-7}
& Tampering 
& Firmware integrity 
& Firmware Exploitation~\cite{cuozzo2016critical} 
& Forged firmware updates with malicious code 
& Secure boot, firmware signing, and verification 
& Vendors \\ \cline{2-7}
& Repudiation 
& Traceability of changes 
& Firmware Exploitation~\cite{cuozzo2016critical} 
& Unauthorized firmware updates to deny operations 
& Immutable logs and event tracking systems 
& Implementers \\ \cline{2-7}
& Info. Disclosure 
& Encrypted data breach 
& Cryptanalysis Attacks~\cite{brassard1997quantum,grover1996fast,shor1994algorithms} 
& Advanced cryptanalysis leveraging quantum algorithms 
& Enhance encryption with quantum-safe solutions 
& Vendors \\ \cline{2-7}
& Elevation of Privilege 
& Unauthorized access control 
& Firmware Exploitation~\cite{cuozzo2016critical} 
& Injecting malicious firmware to gain higher privileges 
& Access control mechanisms, multi-factor authentication 
& Users \\ \hline
\multirow{8}{*}{Storage}
& \multirow{2}{*}{Spoofing} 
& Symmetric cryptographic key vulnerability 
& Brute Force Attack~\cite{shen2020lightweight}
& Via Grover's algorithm to brute-force an n-bit symmetric key 
&Enhanced symmetric encryption with longer key sizes, Implementation of cryptographic agility
&Vendors, Implementers \\ \cline{2-7}
& Info. Disclosure
& Vulnerabilities in classic cryptosystems 
& Cryptanalysis Attacks~\cite{brassard1997quantum,grover1996fast,shor1994algorithms} 
& Data exfiltration via breaking classic cryptosystems 
& Early adoption of quantum-resistant encryption methods, Regular security audits
& Vendors, Implementers \\ \cline{2-7}
& Denial of Service
&Susceptibility to quantum-complex encryption 
& Ransomware~\cite{craciun2019trends,kilber2021cybersecurity} & Quantum-enhanced ransomware attacks 
& Enhanced intrusion detection systems, Frequent and secure data backups, Ransomware-specific defenses 
&Users, Implementers \\ \cline{2-7}
&
{Elevation of Privilege} &
Forgery of digital signatures &
Data Theft~\cite{Threatinsiderdef} &
Code injection with forged signatures during updates &
Implementation of transitional digital signature standards, Multi-factor authentication for update processes &
Vendors, Implementers \\ \hline
\multirow{17}{*}{Networking} &
{Spoofing}
& Vulnerable symmetric encryption 
& Brute Force Attack~\cite{idhom2020network} 
& Via Grover's algorithm to brute-force a symmetric cryptographic key 
& Early implementation of longer key lengths; Monitoring and updating cryptographic standards based on the latest quantum computing advancements 
& Vendors and Implementers \\ \cline{2-7} 
 & {Tampering} 
 & Insecure Wi-Fi Protocols 
 & Wi-fi Wardriving~\cite{lounis2020attacks} 
 & Via Shor's and Grover's algorithms to break classic cryptosystems 
 & Upgrade to WPA3 with quantum-resistant features; Regular updates and patches to Wi-Fi protocols 
 & Vendors and Users \\ \cline{3-7}
 & 
 & Physical AP vulnerabilities 
 & Physical AP Attacks~\cite{chakraborty2017minimum,lounis2020attacks} & Via quantum algorithms to compromise APs 
 & Regular security assessments of APs 
 & Implementers and Users\\ \cline{3-7} 
 & & Susceptible cryptographic protocols 
 & Cryptanalysis Attacks~\cite{zhang2018novel} 
 & {Via quantum algorithms to break cryptosystems} 
 & Development and integration of quantum-resistant cryptographic protocols; Regular testing and validation 
 & Vendors \\ \cline{3-7} 
 & 
 & Signature forgery risk 
 & MITM~\cite{chakraborty2017minimum}
 & Forged signature on messages by quantum attacker & Digital signature standards resistant to quantum attacks & Vendors \\ \cline{2-7} 
 & \multirow{1}{*}{Repudiation} & Lack of non-repudiation controls 
 & Message Repudiation Attack~\cite{aldeco2012secure} 
 & Forged signatures using quantum algorithms 
& Integration of robust, quantum-resistant non-repudiation mechanisms in messaging and communication systems 
& Vendors and Implementers \\ \cline{2-7} 
 & {Info. Disclosure} 
 & Encryption weaknesses 
 & Packet Sniffing (Over Encrypted Packets)~\cite{shibly2020threat}
 &
 Data exfiltration by breaking cryptosystems with quantum computing 
 & Adoption of advanced, quantum-resistant encryption techniques; Continuous monitoring and updating of encryption protocols 
 & Vendors and Users \\ \cline{3-7} 
 & 
 & Vulnerable data transmission
 & MITM~\cite{aldeco2012secure}
 & Data interception using advanced quantum algorithms & Implementation of secure, quantum-resistant transmission protocols; Regular security assessments of transmission channels 
 & Vendors \\ \cline{3-7}
 & 
 & Weak cryptanalysis defenses 
 & Cryptanalysis Attacks~\cite{brassard1997quantum,grover1996fast,shor1994algorithms} 
 & 
 Cryptanalysis Attacks $\&$ Data theft through advanced quantum cryptanalysis 
 & Conducting regular security audits; Updating cryptographic algorithms to quantum-resistant versions 
 & Vendors and Implementers \\ \cline{2-7}
 &{Elevation of Privilege} 
 & Compromised update mechanisms 
 & System Sabotage~\cite{Threatinsiderdef} 
 & Code injection with forged signature during software update 
 & Establishing secure and authenticated software update protocols; Continuous monitoring of software update processes for anomalies 
 & Vendors and Implementers \\ \hline 
\end{tabular}%
}
\label{table:Pre-Migration}
\end{table*}

Based on the outlined attack vectors, it is clear that various threats across different infrastructure layers are emerging, highlighting the need for a comprehensive reassessment. These emerging threats appear distinctly across diverse infrastructure layers, challenging traditional security frameworks  and requiring a paradigm shift in both defensive strategies and cryptographic standards.  Each layer, from data storage to network transmission, now faces unique vulnerabilities. This situation necessitates a transformation in our security paradigm, which is particularly important in the emerging era of quantum computing. 
In the subsequent sections, we will examine the specific security challenges faced by each infrastructure layer, discuss these challenges in detail, and propose targeted strategies to effectively address them, ensuring a robust defense against these emerging threats.\\

\begin{enumerate}[wide, font=\itshape, labelwidth=!, labelindent=0pt, label*=\textit{C}.\arabic*.]

\item \textit{Application Layer:} 
Quantum computing introduces significant risks to Application Layer security, particularly in the areas of data protection and user authentication. The potential of quantum computing to break current encryption methods poses a serious threat to the confidentiality and integrity of application data. Furthermore, authentication processes that rely on existing cryptographic standards are at risk of becoming obsolete. To combat these threats, it is imperative to adopt quantum-resistant algorithms and enhance security protocols to protect application data and prevent unauthorized access or alterations.
Vulnerabilities at the Application Layer include weak encryption algorithms, insecure code, and flawed permission models, among others. These vulnerabilities can be exploited through various means, such as brute force attacks~\cite{merhav2019universal}, exploitation of application software~\cite{compastie2020virtualization,hoffman2020web}, and assertion repudiation~\cite{grassi2020digital,aldeco2012secure}. Potential attack vectors include using quantum algorithms to decrypt virtual machines (VMs) during migration, injecting code during software updates, and forging signatures or hashes used in authentication.
To mitigate these risks, the countermeasures suggested involve transitioning to lattice-based or other post-quantum cryptographic standards, strengthening key management practices, adopting secure coding standards, code audits, and implementing post-quantum secure signatures. Additionally, deploying quantum-resistant authentication mechanisms, using hardware security modules for key storage, implementing quantum-safe encryption for data-at-rest and in-transit, and utilizing secure key exchange protocols are recommended.
The responsibility for implementing these countermeasures lies with various stakeholders, including software vendors, security architects, application developers, service providers, end-users, service providers, network administrators, infrastructure operators, and network engineers. These stakeholders must collaborate to ensure the Application Layer is fortified against emerging quantum computing threats.\\

\item \textit{Data Layer:} Advancements in quantum computing, particularly algorithms like Shor's, Grover's, and those by Brassard et al., are poised to subvert traditional cryptographic paradigms that secure cloud-based data storage. The remarkable speed at which quantum computing can decrypt data poses a significant risk to both the confidentiality and integrity of information stored in the cloud~\cite{jangjou2022comprehensive,brassard1997quantum,grover1996fast,shor1994algorithms}. This emerging threat necessitates a critical reassessment of Key Management Systems (KMS) and underscores the need for cryptographic systems that are resistant to quantum attacks.
To address these vulnerabilities, it is imperative to integrate quantum-secure cryptographic primitives to protect against both data breaches and unauthorized alterations of data. The industry must emphasize the urgency of transitioning to post-quantum cryptography (PQC) to maintain robust security standards. Countermeasures such as transitioning to quantum-resistant encryption standards, promoting cryptographic best practices, and implementing regular integrity checks are vital. Additionally, enhancing update mechanisms with quantum-resistant signatures and securing distribution networks with robust authentication protocols are necessary steps to safeguard against these quantum threats.
The responsibility for addressing these challenges is shared among various stakeholders, including industry standards bodies, security architects, data storage providers, security engineers, software vendors, system administrators, cryptographers, and software developers. Each plays a crucial role in ensuring the security of data in the face of quantum computing advancements.\\

 \item \textit{Runtime Layer:} Quantum threats at the Runtime Layer pose significant risks, such as the exposure of sensitive in-memory data and the compromise of execution environments like Java Virtual Machines and .NET runtimes. 
 These advanced quantum attacks could potentially bypass traditional security measures, enabling unauthorized access to critical data, including personal user information and proprietary algorithms during their execution phase. To counteract these threats, it is essential to implement quantum-resistant encryption standards and employ hybrid encryption models~\cite{driscoll-pqt-hybrid-terminology-02}. Additionally, transitioning to quantum-resistant encryption protocols and establishing quantum-secure channels are crucial steps. Regular penetration testing, security updates, and security audits are also vital to strengthen the defenses against quantum decryption capabilities. The responsibility for these countermeasures lies with vendors, security architects, implementers, IT security teams, network administrators, and Chief Information Security Officers (CISOs).\\

\item \textit{Middleware Layer:} 
This layer serves as a crucial communication and data-processing hub for various software services. Its reliance on cryptographic protocols, including those for message queuing, transaction management, and authentication, exposes vulnerabilities to threats such as Cross-VM Side-channel Cache Attacks~\cite{shahzad2015virtualization,anwar2017cross,saeed2020cross} utilizing Cryptanalysis techniques~\cite{brassard1997quantum,grover1996fast,shor1994algorithms}. Advances in quantum computing pose a substantial threat, with the potential to decrypt or alter middleware communications, leading to data leaks, unauthorized changes to transactions, and compromised inter-application messages. 
To mitigate these risks, it is imperative to adopt quantum-resistant cryptographic measures to preserve the integrity and confidentiality of data within the middleware layer. Countermeasures, including implementing stateful hash-based signatures, quantum-safe encrypted channels, and quantum-resistant public key infrastructures, can be effective. The responsibility for these implementations lies with both vendors, who must provide the necessary tools and updates, and implementers, who must integrate and maintain these security measures.\\

 \item \textit{Operating System Layer:} 
The security implications for the Operating System (OS) Layer in the face of quantum computing are profound, with the potential for kernel-level cryptographic breaches significantly increased~\cite{compastie2020virtualization}. Such breaches could result in unauthorized privilege escalation or even total system takeover. The advent of quantum computing emphasizes the urgency for OS security enhancements, particularly in the realm of kernel protection and the adoption of quantum-resistant cryptographic measures. To mitigate these risks, it is imperative to fortify the OS security framework and advance core system defenses to counteract the sophisticated exploitation of OS vulnerabilities by quantum capabilities.
Countermeasures include the implementation of Quantum-Secure Signature Schemes and the transition to Quantum-Resistant Certificates to prevent spoofing attacks. For tampering threats, Secure Kernel Development Techniques and Quantum-Resistant Hardware Security Modules (HSMs) are recommended~\cite{crypto4aHSM}. To address information disclosure, integrating hardware-enforced security measures like Trusted Platform Modules (TPMs) and quantum-resistant hardware features, along with regular kernel integrity checks, are vital. For repudiation, enhancing audit and logging mechanisms, utilizing multi-factor authentication, and implementing quantum-resistant cryptographic solutions for signature verification are key. Lastly, to prevent the elevation of privilege, it is crucial to implement hardware-enforced security measures unaffected by quantum decryption capabilities and to conduct regular kernel integrity checks using cryptographic hashes based on post-quantum algorithms. These strategies underscore the collective responsibility of OS developers, system administrators, software vendors, security teams, hardware providers, IT security teams, end-users, cryptographers, security architects, and hardware manufacturers in safeguarding against quantum computing threats.\\

 \item \textit{Virtualization Layer:} 
In the Virtualization Layer, exposure to quantum computing threats is multifaceted, with significant concerns including the targeting of hypervisors and unauthorized access to virtualized infrastructures. These vulnerabilities could lead to breaches in VM data confidentiality (e.g., data exfiltration during VM migration~\cite{oberheide2008empirical}). Advanced quantum computing attacks exploiting cryptographic weaknesses could heavily burden computational and memory resources, resulting in performance degradation and the potential for Denial of Service (DoS) in cloud environments. The phenomenon of VM escape is particularly alarming as it endangers the isolation and integrity of virtualized environments.
To mitigate these threats, it is imperative to upgrade cryptographic algorithms resilient to quantum computing techniques, capable of withstanding attacks from Grover's algorithm. This involves transitioning to larger key sizes and adopting post-quantum cryptographic standards. Additionally, enhancing hypervisor security is crucial, achievable through hardware-enforced isolation and the use of intrusion detection systems to monitor anomalous activities. For VMs, implementing strict access control, segregation of critical VMs, and continuous monitoring can help detect and respond to side-channel attacks.
Furthermore, secure protocols for VM migration, coupled with end-to-end encryption and regular security audits of migration processes, are essential countermeasures to protect against data exfiltration during VM transfers. These measures require a collaborative effort, with vendors responsible for development and innovation, while implementers must ensure the deployment, configuration, and execution of these security enhancements.\\

\item \textit{Hardware Layer:} The advent of quantum computing poses significant security risks to the hardware layer, particularly in cloud server processors and memory components. These risks are primarily due to vulnerabilities that can be exploited by advanced quantum algorithms, such as Shor's and Grover's, leading to breaches in encrypted data. The table highlights specific threats like quantum key-breaking, which can be exploited through brute force attacks~\cite{lu2019attacking}, and firmware integrity issues~\cite{cuozzo2016critical}  that can lead to unauthorized access and control. To mitigate these threats, it is crucial to implement quantum-resistant cryptographic algorithms and enhance encryption with quantum-safe solutions. Additionally, secure boot processes, firmware signing, and verification, as well as immutable logs and event tracking systems, are recommended countermeasures. The responsibility for these countermeasures primarily lies with vendors, who must ensure the deployment of secure hardware trust anchors to protect data confidentiality and prevent theft.\\

\item \textit{Storage Layer:} Quantum computing significantly threatens storage layer security, especially through information disclosure vulnerabilities.  Attackers can exploit classical cryptosystems, leading to unauthorized data access and exfiltration (e.g., with quantum-enhanced ransomware making encrypted data nearly impossible to decrypt~\cite{craciun2019trends,kilber2021cybersecurity}). The rise of quantum algorithms like Shor's and Grover's poses a real risk to data integrity, as they can break current encryption methods. To counteract these threats, it is imperative to adopt quantum-resistant encryption techniques and conduct regular security audits. Enhanced symmetric encryption with increased key sizes and the implementation of cryptographic agility are also vital countermeasures. These strategies are vital for protecting the confidentiality and integrity of data at rest against quantum threats.\\

\item \textit{Network Layer:}

In the era of quantum computing, the network layer finds itself in a new battlefield of cybersecurity. Existing encryptions, as shown in Figure~\ref{fig:classic-impact}, and protocols~\cite{CFDIR}, which are standardized and commonly used today, are potential targets for future quantum-powered adversaries. These adversaries could exploit these weaknesses to decrypt network traffic, thereby putting private communications at risk. To defend against these advanced threats, a shift towards cryptographic standards that can withstand quantum attacks is imperative, along with the strengthening of overall network defenses. This involves the adoption of longer cryptographic keys, regular revision of encryption methods, and the implementation of transmission protocols designed to resist quantum-based cryptanalysis.
Both vendors and implementers bear the responsibility of incorporating these advanced security measures to integrate cutting-edge security solutions. This integration aims to protect data during its journey through cloud-based services and applications. In essence, fortifying the network layer against quantum threats necessitates an anticipatory stance, involving the development and deployment of encryption capable of defying quantum-level attacks and sophisticated security methodologies. Constant vigilance through routine security checks and the timely application of updates and patches is indispensable to ensure the ongoing confidentiality and integrity of network interactions in the cloud computing landscape.\\


\end{enumerate}

Table~\ref{table:Pre-Migration} presents a comprehensive overview of pre-migration security challenges across various infrastructure layers. This inclusive table serves as a quick reference guide, encapsulating key insights derived from our analysis. The table encompasses the following components: (1) \textit{Layer}, identifying the specific infrastructure layer examined (e.g., Data Layer, Application Layer, etc.). (2) \textit{STRIDE Threat}, enumerating STRIDE threats pertinent to each vulnerability. (3) \textit{Vulnerabilities}, elucidating specific vulnerabilities associated with each layer susceptible to exploitation by identified threats. (4) \textit{Exploited by}, detailing the threats responsible for the vulnerabilities. (5) \textit{Attack Vector}, furnishing examples of how attackers could exploit these vulnerabilities (e.g., buffer overflow attacks, side-channel attacks, etc.). (6) \textit{Countermeasures}, presenting a compilation of proposed solutions and practices to mitigate identified threats and vulnerabilities. (7) \textit{Responsibility}, indicating whether mitigation falls under the purview of vendors, implementers, users, or a combination thereof.

\section{Cybersecurity Implications of Quantum Computing on Digital infrastructure in Post-Migration Era}\label{sec:post}

Building upon the insights discussed in the preceding section, this part of our study shifts focus toward the Post-Migration Phase, where quantum-resistant cryptography begins to reshape the digital security landscape. We thoroughly examine the effects of post-quantum cryptographic systems on the infrastructural layers, aiming to pinpoint and mitigate the potential threats that quantum computing could introduce to both existing and future digital ecosystems, including cloud computing platforms.
In this advanced stage of transition towards a quantum-safe infrastructure, we delve into the specific Cyber Impacts that arise when infrastructures shift to quantum-resistant algorithms,  like those in post-quantum cryptographic systems. Our analysis is particularly critical at this moment, as moving to these advanced algorithms is vital to defend against the enhanced computational capabilities of quantum computers, which are predominantly aimed at breaking encryption. However, it is crucial to understand that the impact of quantum computing isn't limited to cryptographic threats alone.
We then navigate through the landscape of quantum computing threats in the Post-Migration Era, scrutinizing the effectiveness, adaptability, and resilience of post-quantum cryptographic systems. Despite their robust design against quantum and classical computational threats, vulnerabilities remain. Quantum attackers are continually probing for weaknesses, leveraging sophisticated quantum technologies in techniques like cryptanalysis, side-channel attacks, code injection, and more. This section emphasizes the need for ongoing vigilance and regular reassessment of vulnerabilities within infrastructures, ensuring consistent security even after the adoption of quantum-resistant cryptographic solutions.

\subsection{Cyber Impact of Quantum Computing in the Post-Migration Phase}

The shift to quantum-safe cryptographic algorithms marks a pivotal step in safeguarding organizations from the advanced computational power of quantum computers. This transition, primarily aimed at preventing quantum computers' ability to break traditional encryption, also brings to the fore new cybersecurity challenges that go beyond mere decryption threats. Key among these are issues related to (a) increased key sizes and network traffic, (b) implementation complexity, (c) performance overheads, and (d) adapting network security devices.

\begin{enumerate}[label=(\alph*),font=\bfseries]
\item \textbf{Increased Key Sizes and Network Traffic:} The adoption of PQC typically results in larger cryptographic keys and ciphertexts. This increase can lead to more fragmented network traffic, posing challenges for systems with limited capabilities in managing and reassembling fragmented data. Consequently, the attack surface for threats associated with data fragmentation broadens.

\item \textbf{Implementation Complexity:} PQC integration adds layers of complexity to existing systems, potentially birthing new security vulnerabilities. These vulnerabilities are particularly pronounced in the management of fragmented network packets, necessitating proactive measures for robust security maintenance in the quantum era.

\item \textbf{Performance Overheads:} PQC algorithms can degrade performance, especially in high-traffic environments. This degradation can strain network infrastructure, heightening susceptibility to attacks exploiting fragmentation vulnerabilities.

\item \textbf{Adapting Network Security Devices:} The increasing volume of PQC-encrypted traffic necessitates the advancement of network security appliances,  including firewalls and intrusion detection systems, to effectively process and inspect this new type of traffic. During the transition phase, there may be temporary vulnerabilities that could exacerbate the effects of current fragmentation-based attacks and potentially lead to the emergence of new cyber threats.
\end{enumerate}
While the shift to Post-Quantum Cryptography (PQC) is essential in mitigating threats from quantum computing, it introduces a range of new cybersecurity challenges. These include performance issues in high-traffic environments, increased vulnerability to DoS attacks due to heightened computational demands, and potential weaknesses in cryptographic protocols exploitable by quantum techniques. Moreover, quantum-safe cryptography, though resistant to decryption, does not inherently safeguard against social engineering attacks like phishing. The transition to PQC also raises risks such as buffer overflow attacks in older systems not designed for larger cryptographic keys and ciphertexts, which could lead to advanced malware injections or the forging of cryptographic signatures. Therefore, adopting PQC is a critical but not all-encompassing solution. A comprehensive security strategy is imperative, one that fortifies systems against a variety of sophisticated threats that could emerge in the post-quantum landscape. The following section will explore various attack vectors that might arise in infrastructures transitioning to PQC.

\subsection{Quantum Computing Attack Vectors in the Post-Migration Phase}

Post-quantum cryptographic systems utilize quantum-resistant algorithms, safeguarding against attacks from both quantum and classical computers. These systems are designed to integrate seamlessly with existing communication protocols and networks after migration. Despite this, quantum attackers persist in their efforts to breach post-quantum cryptography by identifying and exploiting vulnerabilities. They continuously use quantum machines to assess weak spots, applying quantum algorithms, side-channel attacks (as detailed in Table~\ref{tab:side}), cryptanalysis, code injection, and more. This section reviews security challenges may present in infrastructure following the transition to PQ cryptography standards (summarized in Table~\ref{table:Post-Migration}) and explores the associated threats, vulnerabilities, attack vectors, and cryptography types. This information aims to aid security analysts in quickly identifying and mitigating security issues. Below, we discuss potential attack vectors stemming from quantum machines in systems based on PQ cryptography.\\

\begin{enumerate}[wide, font=\itshape, labelwidth=!, labelindent=0pt, label*=\textit{B}.\arabic*.]

\item \textit{Side-Channels:} Quantum attackers are capable of compromising cryptographic algorithms by exploiting implementation weaknesses. Side-channel attacks may stem from the co-location of virtual machines (VMs) or hosts, shared network infrastructure, and other scenarios involving shared resources. Key vulnerabilities include OS Kernel Exploitation~\cite{compastie2020virtualization}, Hypervisor Exploitation~\cite{compastie2020virtualization}, VM Monitoring~\cite{compastie2020virtualization}, Inter-VM Communication Inspection~\cite{compastie2020virtualization}, as well as Cross-VM~\cite{shahzad2015virtualization,anwar2017cross,saeed2020cross}, Micro-architectural Cache~\cite{shen2021micro}, and Fault attacks~\cite{blomer2014tampering}. These vulnerabilities can lead to the disclosure of sensitive information, data tampering, and challenges in ensuring non-repudiation.\\

\item \textit{Code Injection:} 
Quantum attackers might perform code injection, exploiting memory buffer limits to embed malicious or vulnerable code.
In such attacks, malicious or vulnerable code is inserted into a system by exploiting these vulnerabilities, potentially affecting Application Software~\cite{compastie2020virtualization,hoffman2020web}, Hypervisors~\cite{compastie2020virtualization}, and Firmware~\cite{cuozzo2016critical}. Threats may include insecure Code~\cite{owasp_code_injection,hoffman2020web} and Command Injection~\cite{kim2022threat,hoffman2020web}, leading to the execution of Keyloggers~\cite{commomtypmalware}, Viruses~\cite{ThreatsProtection}, Worms~\cite{rawal2023malware}, Trap Doors~\cite{TrapdoorVIRUS}, Fileless Malware~\cite{commomtypmalware}, Trojans~\cite{khan2017stride}, Rootkits~\cite{rawal2023malware}, Spyware~\cite{khan2017stride,commomtypmalware}, Crimeware~\cite{khan2017stride}, VM Escapes~\cite{compastie2020virtualization}, Data Theft~\cite{Threatinsiderdef}, and System Sabotage~\cite{Threatinsiderdef}. Furthermore, attackers might hijack the control flow of applications via buffer-overflow vulnerabilities, emphasizing the critical need for proper memory allocation and validation of key sizes.\\

\item \textit{File System Exploitation:} In the context of post-quantum cryptography and the advancements in quantum computing, file systems and hard drives face heightened risks. These vulnerabilities often stem from network daemons, email clients, or web browsers, which can serve as paths for quantum attackers.  Utilizing buffer overflow vulnerabilities~\cite{masdari2016survey,compastie2020virtualization}, attackers are able to inject wiper malware into systems~\cite{commomtypmalware}. This form of malware is particularly insidious, targeting crucial data for deletion or corruption, leading to potentially severe and irreparable system disruptions.\\

\item \textit{OS Kernel Exploitation:} This method typically involves gaining system control through methods like side-channel attacks, pose risks such as tampering, information disclosure, privilege elevation, rootkit installation, memory corruption, unauthorized data access, and Denial-of-Service (DoS) attacks. While PQC aims to provide robust encryption in the era of quantum computing, its integration into OS kernels needs careful attention. The increased computational and memory requirements of some PQC algorithms could inadvertently introduce new vulnerabilities or exacerbate existing ones, such as buffer overflows~\cite{masdari2016survey,compastie2020virtualization} or timing side-channel attacks~\cite{mushtaq2018run}. These vulnerabilities could allow attackers to bypass security controls, modify system processes, access secure data, and gain administrative privileges. Kernel exploits are particularly dangerous due to their high level of access, with the potential to completely undermine system integrity and confidentiality~\cite{compastie2020virtualization}. Therefore, in addition to adopting PQC, operating systems require rigorous testing and validation of their cryptographic implementations, along with a continued focus on securing against both traditional and quantum-enhanced threats to kernel security.\\

\item \textit{Hypervisor Exploitation:} As organizations transition to post-quantum cryptography, new potential vulnerabilities may emerge in the realm of virtualization, particularly within hypervisor environments. Attackers could exploit these vulnerabilities, such as those in hypervisor interfaces and configurations~\cite{compastie2020virtualization}, possibly using techniques like buffer overflows~\cite{masdari2016survey,compastie2020virtualization}, exacerbated by the larger key sizes and complex data structures of PQC algorithms. Such exploits could lead to the injection of malicious code into the hypervisor or VMs~\cite{owasp_code_injection,hoffman2020web}. One critical consequence of this could be VM Hopping or Guest Jumping, where attackers gain unauthorized access to other VMs on the same host, breaching isolation mechanisms~\cite{compastie2020virtualization,almutairy2019taxonomy,huang2012security}. This type of attack not only undermines the integrity and confidentiality of the targeted VMs but also poses a significant threat to the entire virtualized infrastructure. Therefore, as hypervisors are updated to support PQC, it is crucial to ensure that their security measures are equally advanced to handle the unique challenges presented by PQC, including thorough testing and hardening of the hypervisor against both conventional and quantum-inspired exploitation techniques.\\

\item \textit{VM/Hypervisor Service Exploitation:} The shift from classical to post-quantum cryptography introduces new dynamics in managing virtual environments. With PQC, larger key sizes and potentially more complex cryptographic operations can increase the processing load on VMs and Hypervisors. This increased load could lead to vulnerabilities in scenarios such as VM/Hypervisor Denial-of-Service (DoS) attacks, where the system is overwhelmed by processing excessive encrypted traffic or cryptographic operations~\cite{compastie2020virtualization,huang2012security,masdari2016survey}. Such a scenario might lead to VM/Hypervisor or Cloud-Internal Denial of Service, wherein critical resources (like CPU time and memory) are excessively consumed, potentially causing system crashes or significant performance degradation.
Moreover, the integration of PQC algorithms into VMs and hypervisors must be done with careful consideration to avoid creating new vulnerabilities. For instance, the computational and memory overhead associated with PQC might be exploited in VM Poaching attacks~\cite{almutairy2019taxonomy,kedia2013survey}, where an attacker aims to monopolize the resources of shared VMs, leading to service degradation for other users. It is crucial to monitor and optimally allocate resources in virtualized environments to ensure that the additional demands of PQC do not compromise overall system stability and security. Effective resource management strategies and regular security assessments are key in adapting to the heightened requirements of PQC in virtualized systems.\\

\item \textit{VM Migration Exploitation:} VM migration~\cite{oberheide2008empirical}, the process of transferring a virtual machine between host systems, often involves moving large volumes of encrypted data. Employing post-quantum cryptography algorithms as encryption methods in VM migration processes offers heightened security. However, these algorithms often come with increased computational overhead, potentially impacting the performance and latency of VM migrations. Beyond merely selecting robust PQC methods, there's a need for a well-considered strategy that accounts for the practical implications of their implementation. This strategy must safeguard against quantum-specific threats, such as side-channel attacks tailored to PQC, while also ensuring that VM migrations remain operationally efficient. Thus, the transition to PQC in the realm of VM migration demands a holistic approach, one that prioritizes both quantum-resilient security and practical performance in the evolving cybersecurity landscape.\\

\item \textit{Oversized Cryptography Exploitation:} The adoption of post-quantum cryptography introduces new challenges due to the inherently larger key sizes and, in some cases, larger ciphertexts or digital signatures compared to classical cryptographic algorithms. When these larger cryptographic elements are embedded within messages or data packets, the resulting oversized payloads can strain the processing capabilities of web services or network infrastructure. This strain may lead to increased latency or, in more severe cases, result in a Denial-of-Service (DoS) attack as systems struggle to handle the voluminous or complex encrypted data~\cite{gruschka2006protecting,masdari2016survey}. Careful implementation and optimization of PQC algorithms are therefore necessary to balance the security benefits of quantum resistance against the practical implications of increased data sizes, ensuring that network and service performance is not adversely impacted.\\

\item \textit{Program Stack Exploitation:} In the context of transitioning to post-quantum cryptography, there are additional considerations for the security of the program stack, particularly regarding how stack memory and buffer handling might adapt to the larger key sizes and potentially different performance characteristics of PQC algorithms. While PQC is aimed at securing communications against quantum computer-based threats, its implementation might inadvertently affect how the software manages memory and handles errors, potentially impacting the stack's vulnerability to exploits like Return Oriented Programming (ROP)~\cite{wang2016sigdrop,poulios2015ropinjector}. ROP is an advanced technique that allows attackers to manipulate the program stack to redirect software control flow and execute arbitrary code, usually by taking advantage of buffer overflows and other memory corruption vulnerabilities. The integration of PQC could alter the attack surface for such techniques if not properly managed. It is crucial for developers to be aware of these changes and proactively address potential new vulnerabilities in the program stack that could arise from PQC implementation, ensuring both cryptographic robustness and overall application security against techniques like ROP.\\

\item \textit{Cryptanalysis:} The implementation of PQC algorithms is not immune to cryptanalytic attacks~\cite{zhang2018novel,otmani2010cryptanalysis}. Quantum attackers might exploit implementation flaws or leverage quantum algorithms to devise new types of cryptanalytic attacks, which are not yet fully understood in the current cryptographic landscape. This includes exploiting side-channel vulnerabilities, as detailed in Table~\ref{tab:side}, along with other attack vectors that could be more effectively executed using quantum computers. For instance, Grover's algorithm could be used to enhance brute-force attacks on symmetric cryptographic algorithms, necessitating the doubling of key sizes to maintain equivalent levels of security.
Consequently, cryptanalysis in the quantum era demands not only algorithmic changes but also an increased emphasis on implementation robustness, algorithmic resilience, and the continual assessment of cryptographic strength against emerging quantum techniques. Vigilance in the design and deployment of cryptographic systems becomes paramount in ensuring the confidentiality and integrity of information in a post-quantum world.\\

\item \textit{infrastructure Resource Exhaustion:} The implementation of post-quantum cryptography (PQC) algorithms could significantly impact infrastructure resources. PQC algorithms, designed to be secure against quantum computer-based attacks, often require more computational power and memory than many traditional cryptographic algorithms. This increase in resource demands could lead to scenarios where system infrastructure, including servers and networks, faces higher loads, particularly during intensive encryption/decryption operations~\cite{gruschka2006protecting, masdari2016survey}. Such a situation might not only degrade system performance but also increase vulnerability to conventional attack strategies like Denial-of-Service (DoS), where an attacker deliberately overwhelms a system with excessive requests~\cite{compastie2020virtualization, huang2012security, masdari2016survey}. The risk is particularly acute in environments with legacy systems or limited computational resources, which might struggle to cope with the added demands of PQC. Thus, transitioning to PQC necessitates a careful evaluation and potential upgrade of existing infrastructure to ensure it can handle the new cryptographic workload without compromising performance or security.\\
\end{enumerate}

\subsection{Security Challenges in each Infrastructure Layer in the Post-Migration Phase}

Based on the detailed attack vectors associated with quantum machines in systems that have migrated to post-quantum  cryptography, we can discern an array of emerging threats impacting multiple layers of infrastructure. This new landscape necessitates a comprehensive reevaluation of security strategies and measures across these layers:\\

\begin{table*}[!htbp]
\caption{Post-Migration Security Challenges for Each infrastructure Layer}
\large
\renewcommand{\arraystretch}{1.2}
\resizebox{\linewidth}{!}{%
\begin{tabular}{|p{0.15\linewidth}|p{0.17\linewidth}|p{0.32\linewidth}|p{0.35\linewidth}|p{0.52\linewidth}|p{0.59
\linewidth}|p{0.25\linewidth}|}
\hline 
\textbf{Layer} & \textbf{STRIDE Threat} & \textbf{Exploited Vulnerabilities} & \textbf{Exploited by} & \textbf{Attack Vector} & \textbf{Countermeasures} & \textbf{Responsibility} \\ \hline 
\multirow{30}{*}{Application} 
& Tampering 
& Buffer overflow vulnerabilities 
& Application Software Exploitation~\cite{compastie2020virtualization,hoffman2020web} 
& Code injection exploiting memory buffer limits and side-channel attacks as detailed in Table~\ref{tab:side}. 
& Use memory-safe programming languages, maintain up-to-date software patches, and deploy advanced intrusion detection systems 
& Implementers, Vendors \\ \cline{3-7} 
 & 
 & Hypervisor's security gaps 
 & VM Hopping / Guest Jumping~\cite{compastie2020virtualization,almutairy2019taxonomy,huang2012security} 
 & Exploitation of hypervisor vulnerabilities through interface or configuration weaknesses (e.g., buffer overflow and code injection). 
 & Conduct rigorous hypervisor security testing and enforce strict access controls 
 & Vendors, Implementers \\ \cline{3-7}
 & 
 & Insufficient isolation mechanisms 
 & VM Escape to VM~\cite{compastie2020virtualization}
 & Exploiting buffer overflow vulnerabilities for unauthorized memory access 
 & Enhance secure coding standards, monitor for anomalies, and verify isolation protocols 
 & Implementers \\ \cline{3-7} 
 & 
 & Insecure network endpoints 
 & Wiper Malware Injection~\cite{commomtypmalware} 
 & Utilization of buffer overflow vulnerabilities through network daemons, email clients, or web browsers 
 & Deploy comprehensive anti-malware solutions, network firewalls, and promote secure user practices 
 & Users, Vendors \\ \cline{2-7}
& Repudiation
 & Insecure cryptographic implementations leading to stack-based or heap-based buffer overflows 
 & VM Monitoring from VM~\cite{compastie2020virtualization} & Side-channel attacks exploiting VM co-residence on shared resources 
 & Implement secure coding practices such as bounds checking and regular code audits to prevent side-channel vulnerabilities & Vendors, Implementers \\ \cline{3-7}
 & 
& Co-residence of VMs leading to shared resource exploitation & VM Monitoring from Host~\cite{compastie2020virtualization} & Side-channel attacks exploiting shared resources in co-hosted VMs & Employ robust VM isolation mechanisms and strict resource allocation policies & Implementers, Users \\ \cline{2-7}

& Info. Disclosure 
& Buffer overflows, Side-channel attacks, Insecure code execution
&Application Software Exploitation~\cite{compastie2020virtualization,hoffman2020web}, VM Monitoring from VM~\cite{compastie2020virtualization}, VM Monitoring from Host~\cite{compastie2020virtualization}
&Code injection, Side-channel attacks
& Input validation, Secure memory management, Regular software updates, Intrusion detection systems
& Vendors, Implementers \\ \cline{4-7} 
 & 
 & 
 & Inter-VM Communication Introspection~\cite{compastie2020virtualization},Injecting Spyware~\cite{rawal2023malware} 
 & Buffer overflow, Side-channel attacks
 & Encryption of inter-VM communications, Antivirus software, Application whitelisting
 & Implementers \\ \cline{4-7} 
 &
 & 
 &Injecting Worm~\cite{rawal2023malware}, Injecting Rootkit~\cite{rawal2023malware} & Exploiting control flow, Memory corruption
 & Control flow integrity, Memory-safe languages & Implementers \\ \cline{4-7} 
 & 
 & 
 & Side-channel / Math. Analysis Attacks~\cite{chen2010side-channel} 
 & Cryptographic attacks, Math analysis 
 & Post-quantum algorithms, Secure cryptographic implementations 
 & Vendors, Academia, Research community \\ \cline{2-7} 
& Denial of Service 
& Oversized cryptographic payloads 
& Oversized Cryptography~\cite{gruschka2006protecting,masdari2016survey} & Strained processing capabilities due to encrypted or digitally signed oversized payloads 
& Optimized input validation, efficient memory management, robust network infrastructure & Implementers \\ \cline{2-7} 
& Elevation of Privilege 
& Side-Channel Vulnerabilities 
& Data Theft~\cite{Threatinsiderdef} 
& Exploiting buffer overflow or side-channel vulnerabilities for code injection 
& Security audits, updates, and enhanced encryption protocols & Vendors, Implementers 
 \\ \hline
\multirow{8}{*}{Data} 
 & Info. Disclosure 
 & Vulnerability to side-channel attacks and buffer overflows & Injecting Spyware~\cite{rawal2023malware,commomtypmalware}, Injecting Keylogger~\cite{commomtypmalware}, Side-channel / Math. Analysis Attacks including Optical Attack~\cite{roy2022self,joy2011side} 
 &Code injection via buffer overflow, Information leakage through side-channel attacks
 & Implementing quantum-resistant algorithms, Employing memory protection techniques, Regular security audits and patching
 & Vendors (for providing secure hardware/software), Implementers (for secure system configuration), Users (for following best practices) \\ \cline{2-7}
 
 & Denial of Service 
 & Increased computational load due to oversized cryptographic elements 
 & Oversized Cryptography~\cite{gruschka2006protecting,masdari2016survey} & Oversized payload causing system strain 
 & Optimize cryptographic algorithms to reduce size and computational overhead 
 & Implementers \\ \cline{2-7} 
 & Elevation of Privilege 
 & Buffer overflow vulnerabilities 
 & Data Theft~\cite{Threatinsiderdef} 
 & Code injection via buffer overflow 
 & Buffer overflow protections, secure coding practices
 & Vendors and Implementers \\ \hline

\multirow{40}{*}{Runtime} 
& Spoofing 
& Out-of-bounds memory access 
& Command Injections~\cite{kim2022threat,hoffman2020web}
&Buffer overflow attacks
&Implement input validation, use memory-safe languages, enforce least privilege 
& Implementers, Vendors \\ \cline{2-7} 

 & Tampering 
 & Memory corruption vulnerabilities 
 & Application Software Exploitation~\cite{compastie2020virtualization,hoffman2020web} 
 & Code injection (via malware or out-of-bounds memory access vulnerability through buffer overflow) or side-channel attacks as mentioned in Table~\ref{tab:side}. 
 & Use of memory-safe programming languages, regular patch management, and employing runtime protection mechanisms
 & Implementers, Vendors \\ \cline{3-7} 
 
 & 
 & Out-of-bounds memory access 
 & VM Escape to VM~\cite{compastie2020virtualization} 
 & Exploiting buffer overflow or code injection that executed at the root level 
 & Enhanced VM monitoring, isolation techniques, and vulnerability scanning 
 & Vendors, Implementers \\ \cline{3-7} 
 
 & 
 & Program stack integrity compromise 
 & Return Oriented Programming (ROP)~\cite{wang2016sigdrop,poulios2015ropinjector} 
 & Buffer overflow leading to program stack control and corruption
 & Implementation of control-flow integrity measures and stack canaries 
 & Implementers, Users \\ \cline{3-7} 
 
 & 
 & Injection flaws 
 & Command Injections~\cite{kim2022threat,hoffman2020web} 
 & Exploiting out-of-bounds memory access vulnerabilities through buffer overflow 
 & Input validation, parameterized queries, and use of least privilege principle 
 & Implementers, Users \\ \cline{3-7}
 & 
 & Inadequate input validation 
 & Command Injections~\cite{kim2022threat,hoffman2020web} 
 & Buffer overflow due to out-of-bounds memory access. 
 & Input validation and use of parameterized queries. 
 & Users / Implementers \\ \cline{2-7}

 & Repudiation 
 & Inadequate Logging 
 & Application Software Exploitation~\cite{compastie2020virtualization,hoffman2020web} 
 & Code injection via buffer overflow or side-channel attacks & Implementing comprehensive logging and monitoring systems; Utilizing secure coding practices to prevent buffer overflows 
 & Vendors and Implementers \\ \cline{2-7} 
 
 & Info. Disclosure 
 & Side-Channel Leaks 
 &VM Monitoring from VM~\cite{compastie2020virtualization} 
 & Side-channel attacks exploiting the co-residency of VMs, leading to unauthorized monitoring and data leakage. 
 & Implement robust logging and monitoring systems; Use of hardware-level isolation features. 
 & Vendors, Implementers \\ \cline{4-7} 
 & 
& 
& VM Monitoring from Host~\cite{compastie2020virtualization} & Exploitation of side-channels mentioned in Table~\ref{tab:side} to gain sensitive information from VMs. & Side-channel mitigation techniques; Secure coding practices; Regular security patches. 
& Vendors, Implementers \\ \cline{4-7} 
& 
& 
& Side-channel / Math. Analysis Attacks including Timing attack~\cite{mushtaq2018run}
& Overloading systems through resource-intensive operations or attacks.
& Resource usage monitoring, Load balancing, Redundancy planning 
& Implementers, Vendors \\ \cline{3-7}

& 
& \multirow{3}{=}{Vulnerabilities in memory handling, unsafe buffer limits, out-of-bounds memory access} 
& Command Injections~\cite{kim2022threat,hoffman2020web}
& Out-of-bounds memory access 
& Regular security updates; Memory protection techniques 
& Vendors and Users \\ \cline{4-7}

& 
& 
& Application Software Exploitation~\cite{compastie2020virtualization,hoffman2020web} & Code injection through buffer overflow or side-channel attacks (Table~\ref{tab:side}). 
& Implement memory safety practices and secure coding standards. 
& Vendors, Implementers \\ \cline{3-7}

 & 
 & Interception of Inter-VM Communication
 & Inter-VM Communication Introspection~\cite{compastie2020virtualization} 
 & Exploitation of shared networking infrastructure (Table~\ref{tab:side}). 
 & Strong isolation and encryption mechanisms. 
 & Implementers \\ \cline{2-7}

 &

Denial of Service & Inefficient processing of encrypted traffic & VM Denial of Service~\cite{compastie2020virtualization,huang2012security} & Excessive encrypted traffic causing resource exhaustion & Resource allocation strategies, DoS mitigation solutions & Implementers, Vendors \\ \cline{2-7} 

& Elevation of Privilege & Buffer overflow vulnerabilities & Application Software Exploitation~\cite{compastie2020virtualization,hoffman2020web} & Code injection via malware or side-channel attacks (Table~\ref{tab:side}) & Memory protection mechanisms, Secure coding practices & Implementers, Vendors \\ \cline{3-7}

& & Memory access vulnerabilities, Root-level code execution & VM Escape to VM~\cite{compastie2020virtualization} & Out-of-bounds access via buffer overflow or code injection at root level & Input validation, Stack protection techniques, Use of least privilege principle, code execution controls & Implementers \\ \cline{3-7} 

& & Command injection susceptibility & Command Injections~\cite{kim2022threat,hoffman2020web} & Buffer overflow leading to unauthorized command execution & Boundary checks, Sanitization of input data & Implementers, Users \\ \cline{3-7}
 & 
 & Hypervisor's weakness, interface or configuration vulnerability & VM Hopping / Guest Jumping~\cite{compastie2020virtualization,almutairy2019taxonomy,huang2012security} & Exploiting hypervisor vulnerabilities (e.g., via buffer overflow and code injection) & Hypervisor security hardening, intrusion detection systems, and strict access control & Vendors, Implementers \\ \cline{3-7} 
 & & \multirow{1}{\linewidth}{Insecure command execution} & Command Injections~\cite{kim2022threat,hoffman2020web} & Out-of-bounds memory access via buffer overflow & Input validation, secure coding practices & Users \\ \hline
  \multirow{25}{*}{Middleware} 
& Tampering 
& Buffer overflow vulnerabilities 
& Hypervisor Exploitation~\cite{compastie2020virtualization} & Out-of-bounds memory access via buffer overflow, root-level code execution causing data leakage (host/VM to VM) 
& Implement input validation, use of safe libraries and memory management practices 
& Implementers \\ \cline{2-7} 

 & Repudiation & Buffer overflow attacks leading to unauthorized data access or manipulation. & Hypervisor Exploitation~\cite{compastie2020virtualization} & (a) Out-of-bounds memory access vulnerability via buffer overflow or injection of code, which can be executed at the root level and cause data leakage from host/VM to VM or (b) Side-channel attacks as mentioned in Table~\ref{tab:side}. & Implementation of rigorous input validation, adoption of memory-safe programming practices, and deployment of intrusion detection systems capable of identifying such exploitation attempts. & Vendors and Implementers \\ \cline{2-7}

& Info. Disclosure 
& Side-channel attacks 
& VM Escape to Host~\cite{compastie2020virtualization} 
& Exploiting side-channel vulnerabilities for data leakage (host to VM) 
& Utilize hardware that mitigates side-channel risks, enforce strict access controls 
& Vendors, Implementers \\ \cline{3-7}
&  
& Insecure inter-VM communications & Hypervisor Exploitation~\cite{compastie2020virtualization} 
& (a) Buffer overflow or code injection executed at the root level causing data leakage from host/VM to VM. (b) Side-channel attacks as mentioned in Table~\ref{tab:side}.
& Implement robust isolation mechanisms, update hypervisor security policies, and deploy advanced intrusion detection systems. 
& Vendors and Implementers \\ \cline{3-7} 
& 
& Side-channel vulnerabilities 
& Cross-VM Side-channel Attacks including Cache Attacks and Math. Analysis Attacks~\cite{shahzad2015virtualization,anwar2017cross,saeed2020cross} 
& Information leakage through side-channel/math. analysis attacks. 
& Regularly update and patch hypervisors, use noise-generating techniques to obscure side-channels, and employ strict access controls. 
& Vendors and Implementers \\ \cline{3-7} 

& 
& Micro-architectural vulnerabilities 
& Micro-architectural Cache Side-channel Attacks such as Covert-channel Attacks, Rowhammer, and Transient Execution Attacks~\cite{shen2021micro,lou2021survey} 
& Exploitation of micro-architectural weaknesses leading to data leakage. 
& Harden security against specific attacks (e.g., rowhammer), use hardware that resists known vulnerabilities and perform continuous monitoring. 
& Vendors and Implementers \\ \cline{2-7} 
 & Denial of Service 
 & Resource exhaustion 
 & Hypervisor Denial of Service~\cite{masdari2016survey,compastie2020virtualization} & Overloaded hypervisor due to large encrypted traffic 
 & Load balancing, traffic analysis, DoS protection mechanisms 
 & Implementers, Users \\ \cline{3-7} 
 & & Hypervisor configuration flaws & VM Hopping / Guest Jumping~\cite{compastie2020virtualization,almutairy2019taxonomy,huang2012security} & Exploiting hypervisor interface or configuration via buffer overflow & Secure configuration, regular security audits, hypervisor updates & Vendors, Implementers \\ \cline{3-7} 
 & & Overloaded resource allocation 
 & VM Poaching~\cite{almutairy2019taxonomy,kedia2013survey} 
 & Draining hypervisor resources causing crashes 
 & Resource usage monitoring, infrastructure scaling 
 & Implementers \\ \cline{2-7} 
 & Elevation of Privilege 
 & Unauthorized access to root level 
 & VM Escape to Host~\cite{compastie2020virtualization} 
 & Buffer overflow leading to root level access from VM
 & Strict access controls, vulnerability scanning, privilege restrictions 
 & Implementers, Users \\ \hline

 \end{tabular}%
}
\label{table:Post-Migration}
\end{table*}

\addtocounter{table}{-1}

\begin{table*}[!htbp]
\caption{(Cont.) Post-Migration Security Challenges for Each infrastructure Layer}
\large
\renewcommand{\arraystretch}{1.2}
\resizebox{\linewidth}{!}{%
\begin{tabular}{|p{0.15\linewidth}|p{0.17\linewidth}|p{0.32\linewidth}|p{0.35\linewidth}|p{0.52\linewidth}|p{0.59
\linewidth}|p{0.25\linewidth}|}
\hline 
\textbf{Layer} & \textbf{STRIDE Threat} & \textbf{Exploited Vulnerabilities} & \textbf{Exploited by} & \textbf{Attack Vector} & \textbf{Countermeasures} & \textbf{Responsibility} \\ \hline 

\multirow{65}{*}{OS} 
& Spoofing 
& Kernel-level security breaches 
& Injecting Virus~\cite{ThreatsProtection} 
& Code injection via buffer overflow or side-channel attacks potentially allowing to forge signatures. 
& Implementing rigorous testing and validation of cryptographic implementations, updating kernel protection mechanisms 
& Vendors and implementers \\ \cline{2-7}
 & Tampering 
 & Unauthorized data access or modification, Code injection vulnerability & Injecting Insecure Code~\cite{owasp_code_injection,hoffman2020web} & Exploiting vulnerabilities like buffer overflow or side-channel attacks to modify system processes & Regular security audits, application of buffer overflow protections like stack canaries, ASLR, DEP & Implementers and users \\ \cline{3-7}
 & & Kernel manipulation, Buffer overflow vulnerabilities & OS Kernel Exploitation~\cite{compastie2020virtualization} & Side-channel attacks, Forged signatures via side-channel vulnerabilities & Enhanced kernel protection, Secure coding practices, Regular security patches & Vendors, Implementers \\ \cline{3-7} 
& & Hypervisor's weakness 
& VM Hopping / Guest Jumping~\cite{compastie2020virtualization,almutairy2019taxonomy,huang2012security} 
& Exploiting interface or configuration vulnerabilities 
& Strong isolation policies, secure hypervisor configurations & Vendors \\ \cline{3-7} 
& 
& Out-of-bounds memory access 
& VM Escape to VM~\cite{compastie2020virtualization}, VM Escape to Host~\cite{compastie2020virtualization} & Buffer overflow or injection of code leading to data leakage 
& Memory protection mechanisms, secure coding practices
& Implementers \\ \cline{3-7} 
& 
& Code injection vulnerability 
& Injecting Virus~\cite{ThreatsProtection}, Trojan Horse~\cite{khan2017stride}, Spyware~\cite{khan2017stride,commomtypmalware},  Keylogger~\cite{commomtypmalware}, Crimeware~\cite{khan2017stride} 
& Buffer overflow or side-channel attacks forging signatures & Law enforcement collaboration, cybersecurity training, signature-based detection, behavior monitoring 
& Users \\ \cline{2-7} 
 & Repudiation & Lack of proper logging mechanisms 
 & Cryptanalysis Attacks~\cite{otmani2010cryptanalysis} 
 & Advanced cryptanalysis exploiting PQ algorithm weaknesses & Immutable logs, Blockchain for integrity, Enhanced algorithm validation & Implementers, Users \\ \cline{2-7}

& Info. Disclosure
& Kernel memory exposure 
& OS Kernel Exploitation~\cite{compastie2020virtualization}
& Kernel manipulation via side-channel attacks (i.e., the ones mentioned in Table~\ref{tab:side}).
& Implementing isolation techniques, enhanced access control, and encryption.
& Vendors/Users \\ \cline{3-7}

&& Hypervisor monitoring vulnerabilities 
& VM Monitoring from Host~\cite{compastie2020virtualization} 
& Side-channel attacks mentioned in Table~\ref{tab:side}.
& Hypervisor security hardening, strict VM isolation protocols.
& Vendors \\ \cline{3-7}

&& Inadequate access controls 
& VM Monitoring from VM~\cite{compastie2020virtualization} 
& Side-channel attacks; VM co-residence
& Enhanced access control mechanisms; Secure VM isolation
& Implementers; Vendors \\ \cline{3-7}

&& Inter-VM communication eavesdropping 
& Inter-VM Communication Introspection~\cite{compastie2020virtualization} 
& Side-channel attacks via co-residence of VMs or hosts, common networking infrastructure.
& Secure communication channels, enhanced VM monitoring.
& Implementers \\ \cline{3-7}

&& Insufficient process isolation 
& VM Monitoring from Host~\cite{compastie2020virtualization} 
& Side-channel attacks
& Use of secure and isolated hypervisors; Hardware-based security modules
& Implementers; Vendors \\ \cline{3-7}

&& Leakage of sensitive information 
& Side-channel attacks including cache attacks~\cite{gruss2019page,esfahani2021enhanced} and software-initiated fault attacks~\cite{krautter2018fpgahammer}
& Information leaked through side-channel/mathematical analysis attacks.
& Side-channel attack mitigation techniques, secure coding.
& Vendors \\ \cline{3-7}

&& Out-of-bounds memory access 
& Injecting Malware (Virus, Spyware, Keylogger)~\cite{ThreatsProtection,commomtypmalware}
& Exploiting buffer overflow vulnerabilities, Side-channel attacks
& Input validation, secure coding practices, regular software updates, application whitelisting, memory-safe languages, use of ASLR and DEP
& Implementers, Vendors \\ \cline{3-7}

&& Unauthorized data access and leakage 
& Injecting Insecure Code~\cite{owasp_code_injection,hoffman2020web}
& Code injection via buffer overflow or side-channel attacks.
& Use of hardware security modules, secure boot processes.
& Vendors \\ \cline{3-7}

&& Sensitive data exposure through malicious software 
& Injecting Trojan Horse~\cite{commomtypmalware}
& Code injection via buffer overflow or side-channel attacks.
& Antivirus software, intrusion detection systems.
& Users \\ \cline{3-7}

&& System compromise leading to data leakage 
& Injecting Fileless Malware~\cite{commomtypmalware}
& Hijacking control flow of applications via ROP/JOP.
& Behavior-based detection systems, system integrity checks.
& Implementers \\ \cline{3-7}

&& Propagation of malware with data extraction capabilities 
& Injecting Worm~\cite{rawal2023malware}
& Hijacking application control flow via buffer-overflow vulnerability.
& Network segmentation, proactive threat hunting.
& Implementers \\ \cline{3-7}

&& Side-channel Vulnerabilities 
& Cross-VM Side-channel Cache Attacks~\cite{shahzad2015virtualization,anwar2017cross,saeed2020cross}
& Leveraging shared cache resources for data leakage
& Isolation techniques, Cache partitioning
& Vendors \\ \cline{2-7} 

 & Denial of Service & Excessive resource consumption by PQC algorithms & infrastructure Resource Exhaustion & Overwhelming cryptographic engines with quantum operations & Scalable infrastructure, Load balancing, Resource management & Vendors, Implementers \\ \cline{3-7}
&& Buffer Overflow Vulnerability & Buffer Overflow~\cite{masdari2016survey,compastie2020virtualization} & Out-of-bounds memory access vulnerability causing system crashes or performance degradation. & Implement stack canaries, ASLR, DEP; Regular memory checks; Code auditing; Input validation & Vendors, Implementers \\ \cline{3-7} 
 & & VM Resource Exhaustion & VM Denial of Service~\cite{masdari2016survey,compastie2020virtualization,huang2012security} & VMs exhausted by huge encrypted traffic, leading to service disruption. & Enhanced VM monitoring; Resource usage limits; Traffic shaping & Implementers, Users \\ \cline{3-7} 
 & & Hypervisor Resource Exhaustion & Hypervisor Denial of Service~\cite{masdari2016survey,compastie2020virtualization} & Hypervisors exhausted by excessive encrypted traffic, leading to service disruption. & Robust access control; Enhanced hypervisor security; Traffic shaping & Vendors, Implementers \\ \cline{3-7} 
 & & Malware Induced System Overload & Injecting Virus~\cite{ThreatsProtection} & Malware through code injection overloading system resources. & Antivirus and anti-malware solutions; System hardening; Regular updates & Vendors, Implementers, Users \\ \cline{3-7} 
 & & System Compromise via Insecure Code & Injecting Insecure Code~\cite{owasp_code_injection,hoffman2020web} & Injection of insecure code leading to unauthorized actions or system overload. & Secure coding practices; Memory-safe programming; Dynamic analysis tools & Implementers, Users \\ \cline{2-7}

 & Elevation of Privilege & Hypervisor vulnerabilities, Interface or configuration issues & VM Hopping / Guest Jumping~\cite{compastie2020virtualization,almutairy2019taxonomy,huang2012security} & Exploiting hypervisor weaknesses (e.g., buffer overflow, code injection) & Hypervisor security hardening, Use of HSMs, Strong isolation policies & Vendors, Implementers \\ \cline{3-7} 
 & & Kernel manipulation vulnerabilities & OS Kernel Exploitation~\cite{compastie2020virtualization} & Side-channel attacks & Enhanced kernel isolation, Secure coding practices, Regular patching & Implementers \\ \cline{3-7} 
 & & Hypervisor vulnerabilities & VM Hopping / Guest Jumping~\cite{compastie2020virtualization,almutairy2019taxonomy,huang2012security} & Exploiting buffer overflow and code injection vulnerabilities & Hypervisor security updates, Intrusion detection systems & Vendors, Implementers \\ \cline{3-7} 
 & & Out-of-bounds memory access & VM Escape to VM~\cite{compastie2020virtualization} & Buffer overflow or code injection attacks & Memory protection mechanisms, Access controls & Vendors, Implementers \\ \cline{3-7} 
 & & Unauthorized root-level execution & VM Escape to Host~\cite{compastie2020virtualization} & Buffer overflow or code injection leading to data leakage & VM monitoring tools, Isolation techniques & Vendors, Implementers \\ \cline{3-7} 
 & & Code execution via vulnerabilities & Injecting Virus~\cite{ThreatsProtection} & Code injection through side-channels & Antivirus software, Security-hardened OS configurations & Users, Implementers \\ \cline{3-7} 
 & & Insecure code execution paths & Injecting Insecure Code~\cite{owasp_code_injection,hoffman2020web} & Code injection through buffer overflow & Secure code reviews, Static and dynamic analysis tools & Implementers \\ \hline
\multirow{10}{*}{Virtualization} 
 & Info. Disclosure 
 & Side-channel susceptibility 
 & Side-channel Attacks including Cache Attack~\cite{irazoqui2015s} 
 & Information leakage via side channels 
 & Implementation of strict isolation policies and hardware security modules 
 & Vendors \\ \cline{3-7}

 & 
 & Susceptibility to side-channel attacks 
 & Cross-VM Side-channel Cache Attacks~\cite{shahzad2015virtualization,anwar2017cross,saeed2020cross} 
 & Information leaked through side-channel / math. analysis attacks (i.e., the ones mentioned in Table~\ref{tab:side}). & Implementation of strict isolation mechanisms and regular security audits to detect and patch vulnerabilities. & Vendors, Implementers \\ \cline{3-7}

 & 
 & Micro-architectural vulnerabilities 
 & Micro-architectural Cache Attacks~\cite{shen2021micro} 
 & Information leaked through micro-architectural side-channel attacks. 
 & Regular patching and use of side-channel resistant algorithms, enhanced monitoring of micro-architectural behavior and anomaly detection systems. 
 & Vendors, Implementers \\ \cline{2-7}

 & Denial of Service 
 & Resource exhaustion 
 & Cloud-Internal DoS (CIDoS)~\cite{masdari2016survey} 
 & Host resources exhausted by heavy encrypted traffic from malicious VMs. 
 & Resource allocation limits and traffic shaping policies to prevent resource exhaustion and  anomaly detection. 
 & Implementers, Users \\ \cline{3-7}

 & 
 & VM migration traffic 
 & VM Migration~\cite{oberheide2008empirical} 
 & Heavy traffic during encrypted VM migration causing service delays. 
 & Optimized VM migration protocols to handle PQC encryption loads efficiently and and load balancing. 
 & Vendors, Implementers \\
\hline 
\multirow{9}{*}{Hardware} 
& Tampering 
& Firmware vulnerabilities, Side-channel attack susceptibility 
& Firmware Exploitation~\cite{cuozzo2016critical} 
& Code injection via buffer overflow, Side-channel attacks (Table~\ref{tab:side}) 
& Firmware integrity checks, Hardware-based security modules, Side-channel attack mitigation techniques 
& Vendors, Implementers \\ \cline{2-7} 

& Repudiation 
& Lack of firmware execution traceability 
& Firmware Exploitation~\cite{cuozzo2016critical} 
& Code injection via buffer overflow, Side-channel attacks (Table~\ref{tab:side}) 
& Secure logging, Non-repudiation mechanisms 
& Implementers, Users \\ \cline{2-7} 

& Info. Disclosure 
& Side-channel data leakage 
& Side-channel Attacks~\cite{schellenberg2018remote, sayakkara2019survey, accikkapi2019side, genkin2017acoustic, le2019algebraic, breier2020countermeasure} 
& Power-monitoring, Electromagnetic, Acoustic cryptanalysis, Differential Fault Analysis attacks (Table~\ref{tab:side}).
& Side-channel resistant hardware design, Isolation techniques, Shielding, Noise generation
& Vendors, Implementers \\ \cline{2-7} 
 \cline{2-7} 

& Denial of Service 
& Resource exhaustion 
& Energy-Oriented DoS~\cite{masdari2016survey} 
& Overburdened traffic causing resource exhaustion 
& Resource management, DoS mitigation protocols 
& Implementers, Users \\ \cline{2-7} 

& Elevation of Privilege 
& Firmware vulnerabilities, Side-channel attack susceptibility 
& Firmware Exploitation~\cite{cuozzo2016critical} 
& Code injection via buffer overflow, Side-channel attacks (Table~\ref{tab:side}) 
& Least privilege principle, User access control, Secure boot processes 
& Vendors, Implementers \\ \hline 

\multirow{8}{*}{Storage} 
& Info. Disclosure 
& Side-channel susceptibility 
& Side-channel Attack~\cite{brumley2015cache} 
& Exploited through cache attacks and co-residency of attacker and victim. 
& Side-channel resistant hardware design, Isolation techniques, Shielding, Noise generation. 
& Vendors and Implementers \\ \cline{3-7}
 & 
 & Data remanence 
 & Data Remanence Attack~\cite{joshi2018standards} 
 & Cold boot attacks performed on memory dump. 
 & Encryption of all sensitive data at rest and secure deletion protocols. 
 & Implementers and Users \\ \cline{2-7}
 & Denial of Service & Resource exhaustion 
 & Energy-Oriented DoS~\cite{masdari2016survey} 
 & Overburdened traffic draining infrastructure resources during encryption/decryption. 
 & Upgrading infrastructure to handle increased loads and DDoS mitigation strategies, optimized cryptographic algorithms, load balancing. 
 & Vendors and Implementers \\ \cline{2-7}
 & Elevation of Privilege 
 & Buffer overflow vulnerabilities 
 & Data Theft~\cite{Threatinsiderdef} 
 & Code injection via out-of-bounds memory access vulnerability through buffer overflow or side-channel attacks. 
 & Buffer overflow protection mechanisms and regular software updates. 
 & Vendors and Implementers \\ \hline 
\multirow{8}{*}{Networking} 
&Tampering 
& Susceptibility to Cryptanalysis 
& Cryptanalysis Attacks~\cite{zhang2018novel} 
& Mathematical analysis attacks mentioned in Table~\ref{tab:side}. 
& Regular algorithm updates, enhanced encryption methods 
& Vendors, Implementers \\ \cline{3-7} 
 & 
 & Vulnerable to Side-channel Leakage 
 & {Side-channel} Attacks~\cite{lounis2020attacks} 
 & Information leaked through side-channel attacks used to alter system processes or data.
 & Enhanced isolation, noise generation techniques 
 & Vendors, Implementers \\ \cline{3-7} 
 & & Risk of Fault Injection 
 & Fault Attacks~\cite{blomer2014tampering} 
 & Information leakage through fault attacks 
 & Fault detection and correction mechanisms 
 & Vendors, Implementers \\ \cline{2-7} 
 & Info. Disclosure 
 & Risk of Data Remanence 
 & {Side-channel} Attacks including Data Remanence Attack~\cite{naik2021network} 
 & Exploited through cold boot attacks performed on memory dump
 & Secure data erasure protocols, encryption at rest & Users, Implementers \\ \cline{2-7} 
& Denial of Service 
& Fragmentation-induced Overhead
& TCP / UDP Fragmentation~\cite{jafarigiv2020scalable,muller2020retrofitting,essay89509} 
& Oversized encrypted messages causing re-transmission 
& Traffic management, buffer size adjustments 
& Vendors, Implementers \\ \cline{2-7} 
 & Elevation of Privilege 
 & System Configuration Exploits 
 & System Sabotage~\cite{Threatinsiderdef} 
 & Code injection via buffer overflow leading to system sabotage 
 & Memory-safe programming, access controls 
 & Users, Implementers \\ \hline

 \end{tabular}%
}
\label{table:Post-Migration}
\end{table*}

\begin{enumerate}[wide, font=\itshape, labelwidth=!, labelindent=0pt, label*=\textit{C}.\arabic*.]

\item \textit{Application Layer:} 
The Application Layer faces significant post-quantum cryptography challenges, particularly in key and ciphertext management. These issues heighten the risk of buffer overflow vulnerabilities~\cite{masdari2016survey,compastie2020virtualization}, which are exploited by various attack vectors, including side-channel attacks~\cite{gruss2019page,esfahani2021enhanced} that leverage VM co-residence~\cite{compastie2020virtualization}. Malicious entities can exploit these vulnerabilities to launch a range of security threats, such as APTs, keyloggers, ransomware, and polymorphic malware, leading to severe consequences like application control flow manipulation and unauthorized data access.
To mitigate these risks, secure coding practices are essential, including bounds checking and code audits to prevent side-channel vulnerabilities. Robust VM isolation and strict resource allocation policies are crucial to protect against shared resource exploitation. Employing memory-safe programming languages, keeping software up-to-date, and deploying advanced intrusion detection systems are critical to prevent tampering and unauthorized access. Additionally, comprehensive defense strategies should include anti-malware solutions, network firewalls, and the promotion of secure user behaviors to safeguard against vulnerable network endpoints.
For information disclosure threats, rigorous input validation, secure memory handling, frequent software patching, and intrusion detection systems are vital. Enhancing security with encrypted inter-VM communications, antivirus software, and application whitelisting provides further protection against spyware and malware.
To address denial of service attacks, optimized input validation, effective memory management, and a resilient network infrastructure are necessary to handle oversized cryptographic payloads. Regular security audits, prompt application of updates, and advanced encryption protocols are imperative to prevent privilege escalation incidents by addressing potential side-channel vulnerabilities.
The responsibility for implementing these security measures lies with vendors and implementers, and to some extent, users themselves.\\

\item \textit{Data Layer:} This layer is susceptible to information disclosure threats, particularly from quantum-capable adversaries who may utilize advanced cryptanalytic techniques to exploit vulnerabilities such as side-channel attacks and buffer overflows. These vulnerabilities can be exploited through various means, including the injection of spyware~\cite{rawal2023malware,commomtypmalware} or keyloggers~\cite{commomtypmalware}, as well as mathematical analysis attacks, such as optical attacks~\cite{roy2022self,joy2011side}. The use of post-quantum cryptographic algorithms, while enhancing security, also introduces challenges due to their larger key and ciphertext sizes. These increased sizes can lead to a higher likelihood of buffer overflow attacks, which not only threaten data confidentiality and integrity but also present challenges for cryptographic agility and backward compatibility. To mitigate these risks, it is essential to implement quantum-resistant algorithms and employ memory protection techniques. Regular security audits and timely patching are also crucial. The responsibility for addressing these concerns is shared among vendors, who must provide secure hardware and software; implementers, who are responsible for secure system configuration; and users, who should adhere to best practices.\\

\item \textit{Runtime Layer:} 
At the runtime layer, adversaries may exploit vulnerabilities such as out-of-bounds memory access or memory corruption to facilitate attacks like Return Oriented Programming (ROP)  and Jump-Oriented Programming (JOP)~\cite{wang2016sigdrop,poulios2015ropinjector}. These sophisticated techniques manipulate the application's control flow by utilizing sequences of existing code in the runtime memory, potentially leading to unauthorized actions or the bypassing of security mechanisms. To counter such threats, it is crucial to implement control-flow integrity measures, stack canaries, and to use memory-safe programming languages. Regular patch management and runtime protection mechanisms are also vital in mitigating these risks. Responsibility for these countermeasures lies with both implementers and vendors, who must ensure the integrity and security of the runtime environment.\\

\item \textit{Middleware Layer:} 
This Layer, serving as a critical bridge between applications and operating systems, is susceptible to a range of security threats. Notably, the misuse of oversized cryptographic payloads poses a significant risk. Such payloads can lead to buffer overflow vulnerabilities, which, if exploited, may result in denial of service attacks, service impairment, or even arbitrary code execution~\cite{masdari2016survey,compastie2020virtualization,almutairy2019taxonomy,kedia2013survey}. These incidents can undermine the trustworthiness and reliability of middleware services.
To mitigate these risks, it is essential to implement input validation and adopt safe libraries and memory management practices. Furthermore, securing hypervisor configurations and interfaces against buffer overflow exploitation is crucial. Regular security audits, updates, and the use of hardware that mitigates side-channel risks can also play a vital role in protecting the middleware layer. Implementing robust isolation mechanisms, updating hypervisor security policies, and deploying advanced intrusion detection systems are additional countermeasures that can help safeguard against unauthorized data access or manipulation, resource exhaustion, and elevation of privilege attacks. The responsibility for these countermeasures falls on both vendors and implementers, highlighting the need for a collaborative approach to middleware security.\\

\item \textit{Operating Systems Layer:} 
The Operating Systems (OS) layer is a critical component that, if compromised, can lead to significant security breaches. This layer is vulnerable to a variety of security threats, including spoofing, tampering, information disclosure, denial of service (DoS), and elevation of privilege. These threats exploit various vulnerabilities, such as kernel-level security breaches~\cite{compastie2020virtualization}, buffer overflow~\cite{masdari2016survey,compastie2020virtualization}, and side-channel attacks~\cite{gruss2019page,esfahani2021enhanced}, which can lead to unauthorized actions like forging signatures, data alteration, and system overloads.
To mitigate spoofing threats, it is essential to conduct thorough testing and validation of cryptographic implementations and to keep kernel protection mechanisms up to date. Tampering, which involves unauthorized data access or modification, can be countered by regular security audits and implementing buffer overflow protections like stack canaries~\cite{zhou2022final}, Address Space Layout Randomization (ASLR)~\cite{nicula2019exploiting} , and Data Execution Prevention (DEP)~\cite{MicrosoftDEP2023}.
Information disclosure risks, such as kernel memory exposure or hypervisor monitoring vulnerabilities, can be addressed through isolation techniques, enhanced access control, encryption, and hypervisor security hardening. DoS attacks, which aim to consume excessive resources or induce system overload, can be mitigated by scalable infrastructure, load balancing, resource management, and robust access control measures.
Elevation of privilege occurs when attackers exploit hypervisor vulnerabilities or out-of-bounds memory access to gain unauthorized root-level execution. Preventative measures include hypervisor security hardening, memory protection mechanisms, and intrusion detection systems.
The responsibility for implementing these countermeasures is shared among vendors, implementers, and users. Secure coding practices, regular patching, cybersecurity training, and the use of antivirus software are fundamental. Additional measures such as immutable logs, blockchain technology for integrity, and enhanced algorithm validation are recommended to prevent repudiation. Collaboration with law enforcement and proactive threat hunting are also crucial in mitigating code injection vulnerabilities. Through these collective efforts, the integrity and security of the OS layer can be maintained against emerging threats.\\

\item \textit{Virtualization Layer:} 
In the virtualization layer, security challenges such as information disclosure and denial of service are prevalent. Hypervisors and virtual machine monitors are vulnerable to side-channel attacks, including cache attacks~\cite{shahzad2015virtualization,anwar2017cross,saeed2020cross}, which can lead to information leakage. These vulnerabilities are exploited through various attack vectors, such as micro-architectural side-channel attacks~\cite{shen2021micro} that can disclose sensitive information. To mitigate these threats, it is essential to implement strict isolation policies, utilize hardware security modules, and conduct regular security audits to detect and patch vulnerabilities. Additionally, resource exhaustion can be caused by cloud-internal denial of service attacks, where host resources are overwhelmed by heavy encrypted traffic from malicious VMs. Countermeasures include setting resource allocation limits, employing traffic shaping policies, and using anomaly detection systems to prevent resource exhaustion. Optimized VM migration protocols are also necessary to handle post-quantum cryptography encryption loads efficiently and maintain service continuity during heavy traffic periods. The responsibility for these countermeasures lies with both vendors and implementers, while users are also responsible for some aspects of denial of service mitigation.\\

\item \textit{Hardware Layer:} 
In the hardware layer, security is compromised by sophisticated side-channel attacks that exploit vulnerabilities in specific components such as cryptographic accelerators and shared caches within multi-core processors. These attacks are particularly concerning in shared or cloud-based infrastructures, as they can jeopardize the security of multiple hardware modules. The vulnerabilities include firmware weaknesses~\cite{cuozzo2016critical} and susceptibility to side-channel attacks~\cite{schellenberg2018remote, sayakkara2019survey, accikkapi2019side, genkin2017acoustic, le2019algebraic, breier2020countermeasure}, which can be exploited through techniques like power-monitoring, electromagnetic, and acoustic cryptanalysis, as well as differential fault analysis attacks.
Countermeasures to these threats involve implementing firmware integrity checks, utilizing hardware-based security modules, and applying side-channel attack mitigation techniques. Additionally, side-channel-resistant hardware design, isolation techniques, shielding, and noise generation can be employed to prevent information disclosure. To address the threat of tampering, secure logging, and non-repudiation mechanisms are recommended. Resource management and DoS mitigation protocols are necessary to prevent denial of service attacks, while the least privilege principle, user access control, and secure boot processes can help prevent the elevation of privilege.
The responsibility for implementing these countermeasures falls on both vendors and implementers, with implementers and users also being accountable for certain aspects such as secure logging and resource management.\\

\item \textit{Storage Layer:} In the storage layer, the integrity and availability of data are at risk due to potential buffer overflow vulnerabilities that can be exploited by sophisticated wiper~\cite{commomtypmalware} and ransomware attacks~\cite{craciun2019trends,kilber2021cybersecurity}. These vulnerabilities can lead to the compromise of various storage systems, including direct-attached storage, network-attached storage, and storage area networks. To mitigate these threats, it is essential to implement buffer overflow protection mechanisms and ensure regular software updates. These countermeasures fall under the responsibility of both vendors and implementers to safeguard against attacks that could damage file systems, RAID configurations, and storage protocols.\\

\item \textit{Network Layer:} 
The network layer is susceptible to a range of vulnerabilities, including those arising from cryptanalysis~\cite{zhang2018novel}, side-channel leakage~\cite{lounis2020attacks}, fault injection~\cite{blomer2014tampering}, and fragmentation~\cite{jafarigiv2020scalable,muller2020retrofitting,essay89509}. These vulnerabilities can lead to tampering, information disclosure, and denial of service.  Sophisticated techniques such as cryptanalysis, side-channel attacks manipulating system processes or data, fault attacks leaking sensitive information, and fragmentation causing service disruptions are commonly employed in these threats.   The risks are particularly pronounced during heavy VM migrations, where oversized encrypted messages can cause re-transmission issues, and the potential for decryption of in-transit data due to weaknesses in network security protocols is a concern. To counter these threats, it is essential to implement regular algorithm updates, enhanced encryption methods, isolation and noise generation techniques, fault detection and correction mechanisms, secure data erasure protocols, traffic management, and buffer size adjustments. Both vendors and implementers are responsible for these countermeasures, with users also playing a role in ensuring encryption at rest and employing memory-safe programming and access controls to prevent system configuration exploits and sabotage.
To counteract these risks, a layered and resilient security approach is essential, incorporating continuous adaptation and quantum-resistant cryptographic integration, alongside robust key management and security architecture practices. This proactive defense should also utilize advanced techniques such as quantum key distribution (QKD), hardware-based entropy sources for random number generation, and continuous security monitoring to detect anomalies suggestive of quantum-level attacks.

\end{enumerate}

Table~\ref{table:Post-Migration} presents a comprehensive synthesis of our analysis regarding post-migration security challenges across various infrastructure layers. Following the structure of Table~\ref{table:Pre-Migration}, this reference distills our extensive review into essential points. It categorizes the content into seven critical columns: (1) \textit{Layer}, which identifies the scrutinized infrastructure component; (2) \textit{STRIDE Threat}, listing applicable STRIDE threats; (3) \textit{Vulnerabilities}, describing specific exploitable weaknesses; (4) \textit{Exploited by}, linking the threats to vulnerabilities; (5) \textit{Attack Vector}, exemplifying exploitation methods; (6) \textit{Countermeasures}, suggesting practical mitigation strategies; and (7) \textit{Responsibility}, designating stakeholder accountability for implementing these countermeasures.

\section{Conclusion}\label{sec:conclusion}

As the digital landscape evolves with the integration of Post-Quantum (PQ) Cryptography, we face not only a technological shift but also a strategic cybersecurity imperative. The emergence of quantum computing brings both unparalleled potential and new vulnerabilities, pressing us to rethink our cryptographic foundations. This paper has highlighted the multifaceted challenges and transformations needed across various infrastructure layers to safeguard against emerging quantum threats.
The transition to PQ cryptography extends beyond merely adopting larger keys and ciphertexts; it necessitates a fundamental reshaping of our security architectures. Each infrastructure layer, from application to network management, must be rigorously examined and fortified in this new era. Addressing these vulnerabilities requires more than algorithmic updates; it demands a holistic and proactive security overhaul.
Our journey towards robust PQ cryptography calls for collaborative efforts across academia, industry, and policy realms. This collective endeavor should not only focus on developing and standardizing resilient PQ algorithms but also on reevaluating our overall security postures to counteract sophisticated quantum-borne threats effectively.
Future research should continue to drive algorithmic innovation, ensuring these new solutions are both secure and practically deployable. Equally important is the concurrent evolution of hardware and software systems to integrate these cryptographic advances smoothly. This progression also involves enhanced threat modeling and interdisciplinary collaboration, crucial for addressing real-world application challenges and elevating public awareness.
In summary, the migration towards PQ cryptography represents a comprehensive, collaborative effort to protect our digital domains against upcoming quantum threats. It underscores the need for ongoing innovation, vigilance, and adaptation in our cybersecurity strategies. As quantum computing continues to advance, it is imperative that our defenses not only keep pace but also anticipate and preempt future vulnerabilities.

\bibliographystyle{IEEEtran}
\bibliography{bibliography}
\end{document}